\documentclass[superscriptaddress,prapplied,amsmath,amssymb,floatfix,reprint,raggedbottom,longbibliography]{revtex4-1}
\usepackage[pdftex]{graphicx}
\usepackage{amsmath}
\usepackage{lipsum}
\usepackage{color}
\usepackage{soul}
\usepackage{balance}
\usepackage{fancyhdr}
\usepackage{booktabs}
\usepackage{enumitem}
\usepackage{listings}
\usepackage{hyperref}
\usepackage[ruled,vlined]{algorithm2e}

\hypersetup{colorlinks=true,linkcolor=blue,citecolor = blue,urlcolor = blue, linktocpage}

\begin{document}
%\linenumbers

\title{Hardware-aware \textit{in situ} learning based on stochastic magnetic tunnel junctions}
\author{Jan Kaiser}
\affiliation{Elmore Family School of Electrical and Computer Engineering, Purdue University, West Lafayette, IN, 47906 USA}
\author{William A. Borders}
\affiliation{Laboratory for Nanoelectronics and Spintronics, Research Institute of Electrical Communication, Tohoku University, Sendai, Japan}
\author{Kerem Y. Camsari}
\email{camsari@ece.ucsb.edu}
\affiliation{Department of Electrical and Computer Engineering, University of California Santa Barbara, Santa Barbara, CA, 93106 USA}
\author{Shunsuke Fukami}
\email{s-fukami@riec.tohoku.ac.jp}
\affiliation{Laboratory for Nanoelectronics and Spintronics, Research Institute of Electrical Communication, Tohoku University, Sendai, Japan}
\affiliation{Center for Innovative Integrated Electronic Systems, Tohoku University, Sendai, Japan.}
\affiliation{Center for Spintronics Research Network, Tohoku University, Sendai, Japan.}
\affiliation{Center for Science and Innovation in Spintronics, Tohoku University, Sendai, Japan.}
\affiliation{WPI-Advanced Institute for Materials Research, Tohoku University, Sendai, Japan.}
\author{Hideo Ohno}
\affiliation{Laboratory for Nanoelectronics and Spintronics, Research Institute of Electrical Communication, Tohoku University, Sendai, Japan}
\affiliation{Center for Innovative Integrated Electronic Systems, Tohoku University, Sendai, Japan.}
\affiliation{Center for Spintronics Research Network, Tohoku University, Sendai, Japan.}
\affiliation{Center for Science and Innovation in Spintronics, Tohoku University, Sendai, Japan.}
\affiliation{WPI-Advanced Institute for Materials Research, Tohoku University, Sendai, Japan.}
\author{Supriyo Datta}
\affiliation{Elmore Family School of Electrical and Computer Engineering, Purdue University, West Lafayette, IN, 47906 USA}
\date{\today}

\begin{abstract}
One of the big challenges of current electronics is the design and implementation of hardware neural networks that perform fast and energy-efficient machine learning. Spintronics is a promising catalyst for this field with the capabilities of nanosecond operation and compatibility with existing microelectronics. Considering large-scale, viable neuromorphic systems however, variability of device properties is a serious concern. In this paper, we show an autonomously operating circuit that performs hardware-aware machine learning utilizing probabilistic neurons built with stochastic magnetic tunnel junctions. We show that \textit{in situ} learning of weights and biases in a Boltzmann machine can counter device-to-device variations and learn the probability distribution of meaningful operations such as a full adder. This scalable autonomously operating learning circuit using spintronics-based neurons could be especially of interest for standalone artificial-intelligence devices capable of fast and efficient learning at the edge.

\end{abstract}
\pacs{}
\maketitle
\section{Introduction}
Conventional computers use deterministic bits to operate and encode information. While this approach is effective for well-defined tasks like arithmetic operations, there are many difficult tasks like stochastic optimization, sampling, and probabilistic inference, which instead are readily addressed by utilizing stochasticity. A promising approach for solving these difficult tasks is using computers that are naturally probabilistic. In a well-known piece, Feynman \cite{feynman_simulating_1982} suggested that in the same way that the use of quantum computers is important to simulate quantum phenomena, a probabilistic computer could be a natural solution to problems that are intrinsically probabilistic. Recently, utilizing spintronics technology, Borders et al. \cite{borders_integer_2019} demonstrated such an autonomously running probabilistic computer consisting of probabilistic bits (\textit{p-}bits) with a stochastic magnetic tunnel junction (\textit{s-}MTJ) which can perform computationally hard tasks like integer factorization.\\
Machine learning is another important field in which probabilistic computation and a large amount of random numbers could be highly beneficial. It holds promise for various tasks like image recognition, medical application and autonomous driving \cite{lecun_deep_2015,esteva_guide_2019,schmidhuber_deep_2015}. For these applications, conventional von Neumann computers are inefficient and alternative computing architectures inspired by information processing in the human brain are of interest \cite{noauthor_big_2018,sze_hardware_2017,grollier_neuromorphic_2020,merolla_million_2014,davies_loihi_2018}. Boltzmann machines offer a promising route for hardware learning due to their local learning rule and tolerance to stochasticity \cite{hinton_training_2002,carreira-perpinan_contrastive_2005,ernoult_using_2019,bojnordi_memristive_2016,nasrin_low_2019,le_roux_representational_2008}. Boltzmann machines are generative stochastic recurrent neural networks having a large application space ranging from optimization to generative machine learning \cite{goos_boltzmann_1987,osborn_fast_1990,salakhutdinov_deep_2009,srivastava_multimodal_2012}. This suggests that building a compact hardware implementation in the form of a probabilistic computer that resembles a Boltzmann machine could be highly beneficial in terms of energy consumption and training speed. While some hardware implementations have been presented for Restricted Boltzmann machines (RBMs) \cite{bojnordi_memristive_2016,eryilmaz_training_2016,tsai_413267_2017}, in this paper we focus on fully-connected unrestricted Boltzmann machines. The usual problem in learning unrestricted Boltzmann machines is that they are hard to train since the equilibrium samples of the network are harder to extract \cite{salakhutdinov_deep_2009,salakhutdinov_learning_2008}. In this work we show a system that performs this sampling naturally and could hence make it possible to train unrestricted Boltzmann machines more efficiently using the natural physics of \textit{s-}MTJs. \\
A common concern for the development of neuromorphic systems based on emerging devices like \textit{s-}MTJs is the inevitable device variability \cite{grollier_neuromorphic_2020,de_rose_variability-aware_2017}. This poses an obstacle to deploy these systems for real-world application on a large scale while preserving high reliability. Several approaches have been proposed to overcome these challenges on a device level for example by applying external magnetic fields \cite{lv_experimental_2019}, performing a calibration phase \cite{borders_integer_2019} or by postprocessing \cite{qu_variation-resilient_2018}. Another interesting approach to counter the effect of variability and realize high performance in neuromorphic systems is to perform training and inference on the same hardware system \cite{li_efficient_2018,dalgaty_situ_2021,kiraly_atomic_2021}. In this paper, we present a proof-of-concept demonstration of a probabilistic computer that can perform \textit{in situ} learning allowing to counter device-to-device variations naturally as part of its learning process. Here, device variability is addressed on a system’s level. We show that devices with nonideal characteristics can be used to perform given tasks successfully without the necessity to individually calibrate each device. This is achieved by learning hardware-aware weights and biases. Such a natural variation tolerance could enable large-scaled implementations of MTJ-based probabilistic computers.\\

\begin{figure*}
    \setlength\abovecaptionskip{-0.5\baselineskip}
    \centering
    \includegraphics[width=1\linewidth]{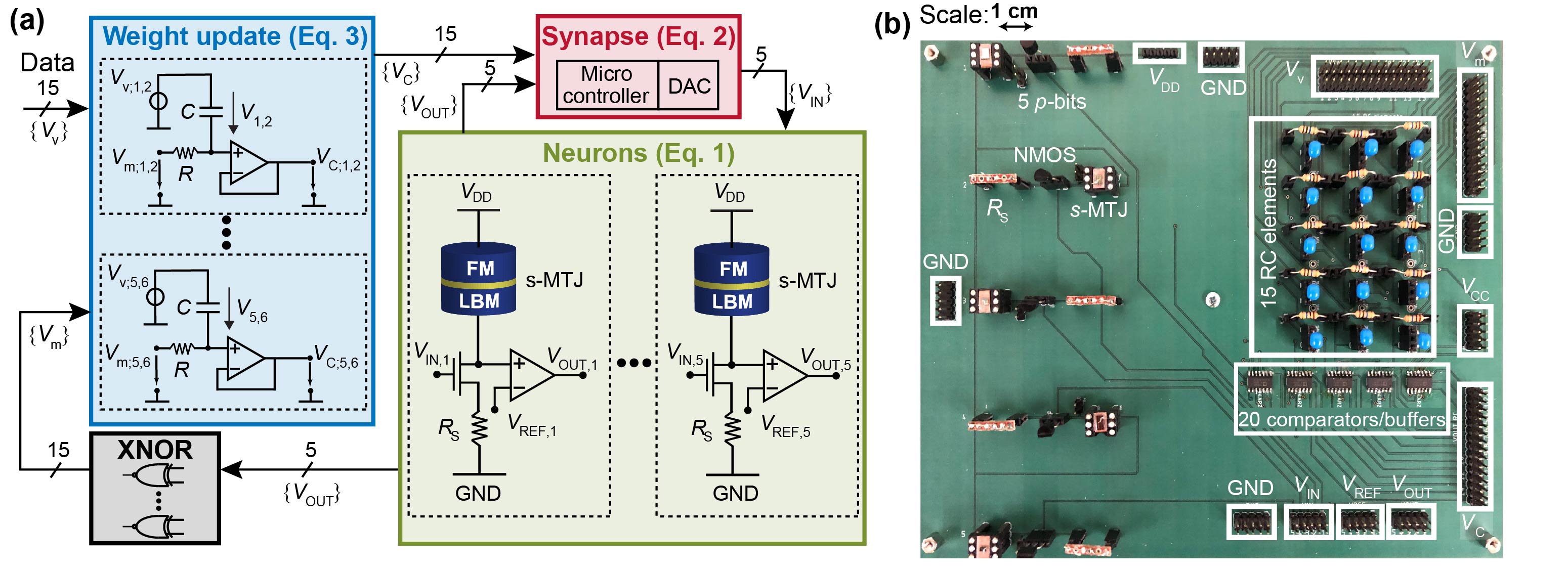} 
    \vspace{0.1in}
    \caption{\textbf{Probabilistic Learning Circuit:} (a)  Block diagram of the learning circuit with \textit{p-}bit output voltages $\{V_{\rm OUT}\}$, \textit{p-}bit input voltages $\{V_{\rm IN}\}$, weight voltages $\{V_{\rm i,j}\}$, capacitor voltages $\{V_C\}$, \textit{p-}bit correlation voltages $\{V_{\rm m}\}$ and data distribution correlations voltages $\{V_{\rm v}\}$. (b)  A photograph of the PCB with the 5 \textit{p-}bits (each consisting of an \textit{s-}MTJ, an NMOS transistor and a source resistor $R_{\rm S}$) and 15 RC elements and 20 operational amplifiers (5 used as a comparator and 15 as a buffer). The \textit{p-}bits are interconnected with the RC-array as shown in (a).}
    \label{fig:circuit}
\end{figure*}

\section{Hardware-aware learning with MTJ-based \textit{p-}bits}
The main building block of a probabilistic computer is the \textit{p-}bit, analogous to a binary stochastic neuron (BSN) \cite{ackley_learning_1985}. Its activation function can be described by \cite{camsari_stochastic_2017}
\begin{equation}
m_{\rm i} (t+\tau_{N}) = {\rm{sgn}}\left[ \tanh \left( {I_{\rm i}(t)}\right) - r\right] .
\label{eq:binary_stochastic_neuron}
\end{equation}
Here, $m_{\rm i}$ is the output of the \textit{p-}bit and a bipolar random variable, $\tau_N$ is the time the \textit{p-}bit takes to perform the activation operation, $I_{\rm i}$ is the dimensionless input to \textit{p-}bit $i$, and $r$ is a uniformly distributed random number between -1 and +1. Eq.\eqref{eq:binary_stochastic_neuron} can also be written in binary notation with a unit step function and a sigmoid function.
To connect multiple \textit{p-}bits, a synaptic function computes the input of every \textit{p-}bit $I_{\rm i}$ by taking the weighted sum of all \textit{p-}bit outputs $m_{\rm i}$,
\begin{equation}
I_{\rm i} (t+\tau_S) =  \sum_{j} W_{\rm i,j} m_{\rm j}(t),
\label{eq:synaptic_function}
\end{equation}
where $\tau_S$ is the synapse execution time and $W_{\rm i,j}$ is the weight matrix that couples \textit{p-}bit $i$ and \textit{p-}bit $j$. Here, the bias to \textit{p-}bit $i$ is subsumed into $W_{\rm i,j}$. Given a particular weight matrix, every \textit{p-}bit configuration has a defined probability given by the Boltzmann distribution where $P(m) \propto \exp{\big[-\beta E(m)\big]}$ with energy $E(m)=-\sum W_{\rm i,j} m_{\rm i} m_{\rm j}$ and inverse temperature $\beta$. For training a Boltzmann machine, the goal is to find a weight matrix W that results in a Boltzmann distribution that fits closely to the given training vectors $\{v\}$. The distribution of training vectors is referred to as data distribution in this paper. To find a fitting weight matrix for a given data distribution, the weights are trained by performing gradient ascent of the log-likelihood \cite{koller_probabilistic_2009}. It is well known that the ideal Boltzmann machine algorithm based on log-likelihood learning is generally intractable since learning time scales exponentially with the size of the system  \cite{nair_implicit_2009,salakhutdinov_deep_2009}. However, it has been shown that approximate version of the Boltzmann learning rule like the contrastive divergence algorithm \cite{hinton_training_2002,carreira-perpinan_contrastive_2005} can be used to perform approximate learning for large Boltzmann machine systems.  This algorithmic scaling motivates the use of domain-specific, efficient, and fast hardware accelerators like the \textit{p-}bit building block that naturally represents the neuron function of the Boltzmann machine in order to accelerate the learning process \cite{hamilton_accelerating_2020}.  
To map the Boltzmann machine learning algorithm to our hardware system, we use a continuous learning rule similar to the persistent contrastive divergence algorithm given by \cite{tieleman_training_2008,kaiser_probabilistic_2020},
\begin{equation}
\frac{dW_{\rm i,j}}{dt} = \frac{\langle v_{\rm i} v_{\rm j} \rangle - m_{\rm i} m_{\rm j} - \lambda W_{\rm i,j}}{\tau_{L}},
\label{eq:learning_rule}
\end{equation}
that can be implemented in hardware. Here, $\langle v_{\rm i} v_{\rm j} \rangle$  is the average correlation between two neurons in the data distribution where $v_{\rm i}$ is the training vector entry for \textit{p-}bit $i$, $m_{\rm i} m_{\rm j}$  is the correlation of the \textit{p-}bit outputs defined in Eq.\eqref{eq:binary_stochastic_neuron} and $\tau_L$ is the learning time constant. Regularization parameterized by $\lambda$ assures that weights do not become too large and helps the algorithm to converge to a solution \cite{ng_feature_2004}. This learning rule requires only the correlation between two \textit{p-}bits $m_{\rm i} m_{\rm j}$  for updating weight $W_{\rm i,j}$ which makes this learning algorithm attractive for hardware implementations. Eq.\eqref{eq:learning_rule} does not change when the system becomes larger. Another advantage of the presented hardware implementation of the Boltzmann machine is that the computational expensive part of getting the equilibrium samples of the correlation term $m_{\rm i} m_{\rm j}$ needed for learning is performed naturally.

Eqs.\eqref{eq:binary_stochastic_neuron},\eqref{eq:synaptic_function},\eqref{eq:learning_rule} are implemented in hardware to build a probabilistic circuit that performs learning. The dimensionless quantities of Eqs.\eqref{eq:binary_stochastic_neuron} and \eqref{eq:synaptic_function} are converted to the physical quantities shown in Fig. \ref{fig:circuit} as follows: $m_{\rm i}=2 \cdot V_{\rm OUT,i}/V_{\rm DD} -1$ with \textit{p-}bit output voltage $V_{\rm OUT,i}$ and $I_{\rm i}=V_{\rm IN,i}/V_0$ with \textit{p-}bit input voltage $V_{\rm IN,i}$ and \textit{p-}bit reference voltage $V_0$ which is defined by the response of the \textit{p-}bit \cite{camsari_implementing_2017}. Eq.\eqref{eq:learning_rule} can be written into circuit parameters using RC elements \cite{kaiser_probabilistic_2020}
\begin{equation}
C \frac{dV_{\rm i,j}}{dt} = \frac{V_{\rm v;i,j} - V_{\rm m;i,j} -V_{\rm i,j}}{R} 
\label{eq:learning_circuit}
\end{equation}
where $V_{\rm i,j}$ is the voltage across capacitor $C$, $R$ is the series resistance, $V_{\rm v;i,j} \ \widehat{=} \ \langle v_{\rm i}v_{\rm j} \rangle$ is the voltage representing the average correlation of two neurons in the data distribution and $V_{\rm m;i,j} \ \widehat{=} \ m_{\rm i}m_{\rm j}$ is the voltage representing the correlation of \textit{p-}bit outputs\footnote{The exact mapping of the correlation voltages $V_{\rm v;i,j}$ and $V_{\rm m;i,j}$ is discussed in the methods section.}. Eqs.\eqref{eq:learning_rule} and \eqref{eq:learning_circuit} can be converted into each other by setting $W_{\rm i,j}=A_v V_{\rm i,j}/V_0$, $\lambda = V_0/(A_v V_{\rm DD}/2)$ and $\tau_{L}=\lambda RC$ where $A_v$ is a voltage gain factor between the voltage across the capacitor and the used weight value for the weighted sum in Eq.\eqref{eq:synaptic_function}. While for memory usage, nonvolatile storage of a capacitor can be detrimental, the discharging of the capacitor is used here as weight decay or regularization in the learning process that ensures that the learning converges. The voltage gain is used to adjust the regularization parameter $\lambda$ for the update rule Eq.\eqref{eq:learning_rule}. High $\lambda$ produces smaller weight values during learning. More information about the learning rule is presented in the supplementary information \footnote{See Supplemental Material at [URL will be inserted by
publisher] for more information regarding the learning rule and learning examples for AND, OR and XOR gates.}. Note that while we choose a RC network in this proof-of-concept experiment to conveniently represent analog voltages as weights, the synaptic functionality in our system could also be implemented out of memristor crossbar arrays \cite{li_efficient_2018,ambrogio_equivalent-accuracy_2018,mahmoodi_versatile_2019} to support \textit{in situ} learning by mapping the weight update rule (Eq.\eqref{eq:learning_rule}) to an equation of changing conductance $G_{\rm i,j}$ instead of changing voltage $V_{\rm i,j}$. The use of memristor crossbars would have the main advantage that the weight storage becomes nonvolatile.\\

 Fig. \ref{fig:circuit} (a) shows the block diagram of the learning circuit. The neurons (Eq.\eqref{eq:binary_stochastic_neuron}) are implemented with an \textit{s-}MTJ in series to a transistor and a resistor $R_{\rm S}$. The random number in Eq.\eqref{eq:binary_stochastic_neuron} is generated by the \textit{s-}MTJ which fluctuates between two resistance values $R_{\rm P}$ and $R_{\rm AP}$ which represents the parallel and anti-parallel configuration of the fixed and free layer of the MTJ. While the fixed layer is a normal ferromagnet (FM), the free layer is designed to be a low-barrier magnet (LBM) which magnetic orientation changes due to thermal noise resulting in resistance fluctuations of the MTJ. The drain voltage gets thresholded by using a comparator \cite{borders_integer_2019,camsari_implementing_2017} where the reference voltage is chosen to be $V_{\rm REF}=V_{\rm DD}-I_{50/50}\big(\frac{R_{\rm P}+R_{\rm AP}}{2}\big)$ with $I_{50/50}$ being the bias current where the stochastic MTJ stays in the parallel and anti-parallel 50\% of the time. The synapse (Eq.\eqref{eq:synaptic_function}) is implemented by using a microcontroller in conjunction with a digital-to-analog converter (DAC) where the \textit{p-}bit output voltages $\{V_{\rm OUT}\}$ and capacitor voltages $\{V_C\}$ with $V_{\rm i,j}  = V_{\rm v;i,j}  -V_{\rm C;i,j}$ are taken as an input. To compute the correlation of \textit{p-}bit outputs $m_{\rm i} m_{\rm j}$ an XNOR gate is needed between the \textit{p-}bit and the learning block (Eq.\eqref{eq:learning_rule}) where the weights are updated using an RC array. Fig. \ref{fig:circuit} (b) shows the printed circuit board (PCB) with the 5 \textit{p-}bits and the RC-array with 15 RC elements used in the experiment. In the methods section (section \ref{subsec: experiment}) more details about the experimental implementation are presented.

\section{Variation-tolerant learning of a full-adder}
We demonstrate the learning of the hardware circuit using the data distribution of a full adder (FA). In general, for a fully visible Boltzmann machine with $N$ \textit{p-}bits, $(N+1)N/2$ weights and biases have to be learned. A FA has 3 inputs and 2 outputs resulting in $N = 5$ \textit{p-}bits. To connect these \textit{p-}bits, 10 weights and 5 biases have to be learned (in total 15 RC elements as shown in Fig. \ref{fig:circuit} (b). For the FA, the binary inputs $[ABC_{\rm in}]$ get added and the outputs are given by the sum $S$ and the carry out $C_{\rm out}$ as shown in Table \ref{table: FA}. This corresponds to a data distribution that is given by 8 out of the 32 ($2^N$) possible configurations. Because of the probabilistic nature of this circuit, input and outputs are treated equally, which allows, for example, invertible full adder operation \cite{camsari_stochastic_2017,camsari_implementing_2017} and distinguishes our probabilistic circuit from conventional logic gates that can operate only in one direction. While we have chosen the FA truth table as data distribution, any probability distribution could be chosen to be represented by our probabilistic circuit. In methods section \ref{subsec: mapping}, the data distribution in form of the truth table of the FA and the mapping from truth table to analog voltages $V_{\rm v;i,j}$ is explained in more detail. For the FA, the learning is performed for a total of 3000 s. In the supplementary information \cite{Note2}, learning examples for an AND, OR and XOR gate with less \textit{p-}bits are shown.\\

\begin{figure*}
    \setlength\abovecaptionskip{-0.5\baselineskip}
    \centering
    \includegraphics[width=1\linewidth]{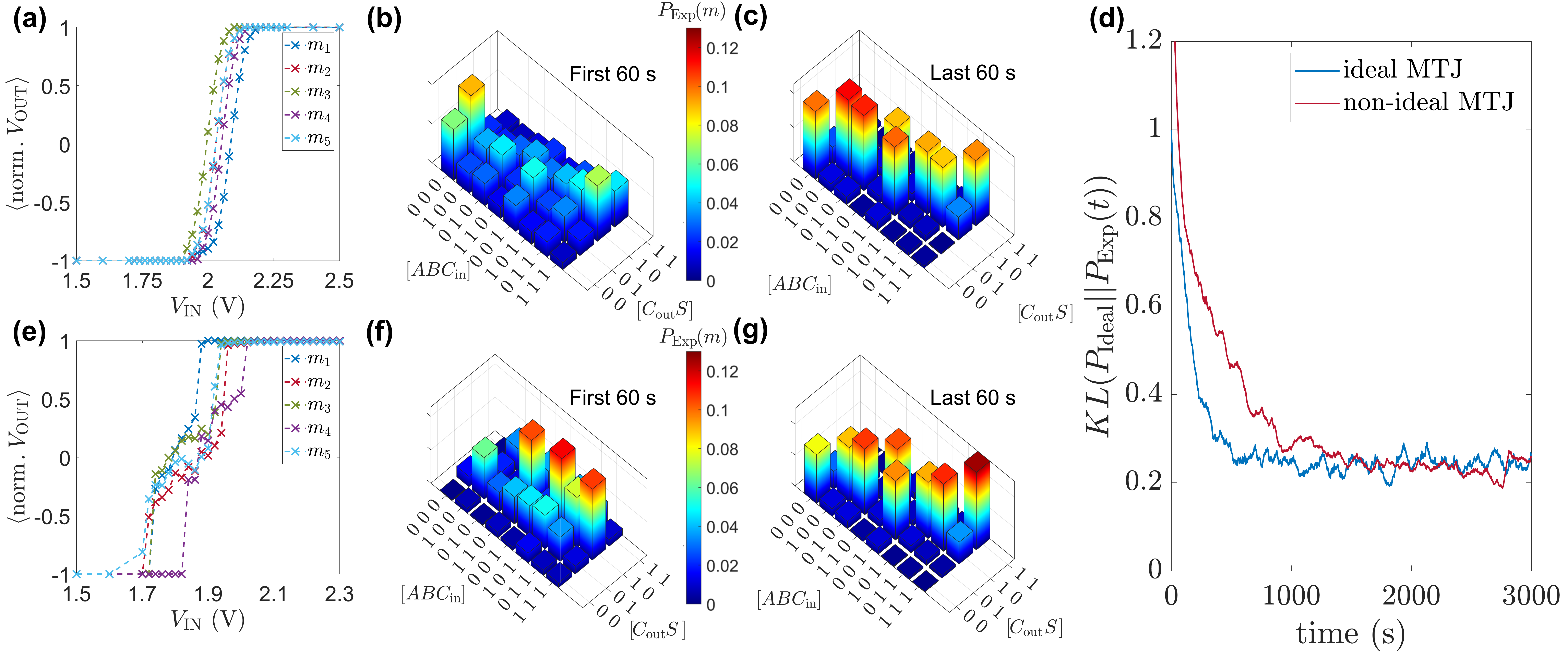} 
    \vspace{0.1in}
    \caption{\textbf{Full Adder (FA) learning:} (a) Average response of emulated ideal MTJ \textit{p-}bits for the 5 \textit{p-}bits used in the FA with average normalized output voltage $\langle \mathrm{norm.} \ V_{\rm OUT,i} \rangle = 2\cdot \langle V_{\rm OUT,i} \rangle /{V_{\rm DD} -1=\langle m_{\rm i} \rangle}$. Every point is averaged over 15 s. (b) Experimental distribution of emulated ideal MTJ circuit $P_{\rm Exp}(m)$ with \textit{p-}bit output states $([m_1,m_2,m_3,m_4,m_5]+1)/2=[A,B,C_{\rm in},S,C_{\rm out}]$ where $m_{\rm i}=2 \cdot V_{\rm OUT,i}/V_{\rm DD} -1$ collected as a histogram over for the first 60 s of learning. (c) Experimental distribution of emulated ideal MTJ circuit collected as a histogram over the last 60 s of learning. (d) KL-divergence between ideal and experimental distribution $KL(P_\mathrm{Ideal} || P_\mathrm{Exp}(t))$ vs. time of ideal and nonideal MTJ system. The experimental distribution is obtained over 60 s of learning. (e) Average response of nonideal MTJ \textit{p-}bits for the 5 \textit{p-}bits used in the FA with average normalized output voltage  $\langle \mathrm{norm.} \ V_{\rm OUT,i} \rangle = 2\cdot \langle V_{\rm OUT,i} \rangle /{V_{\rm DD} -1=\langle m_{\rm i} \rangle}$. Every point is averaged over 15 s. (f) Experimental distribution of nonideal MTJ circuit $P_{\rm Exp}(m)$ collected as a histogram over the first 60 s of learning. (g) Experimental distribution of nonideal MTJ circuit $P_{\rm Exp}(m)$ collected as a histogram over the last 60 s of learning.}
    \label{fig:FA}
\end{figure*}

\subsection{Full adder learning with emulated ideal MTJ}
Fig. \ref{fig:FA} (a) shows the normalized, time averaged \textit{p-}bit response of every \textit{p-}bit using the ideal \textit{s-}MTJ implementation when the input voltage $V_{\rm IN}$ is swept. These \textit{s-}MTJs are emulated in hardware with two resistances that are randomly selected by a multiplexer (MUX) to obtain nearly ideal \textit{p-}bit response characteristics (see methods section \ref{subsec: p-bit} for more details). Due to variations in the circuit, every curve is slightly shifted from the ideal 50/50 point at $V_{\rm IN}=1.95 \ \mathrm{V}$. Even though we are using the MUX model here, it has been shown by Borders et al.\cite{borders_integer_2019} that near ideal \textit{p-}bit responses can be obtained with real \textit{s-}MTJs. In previous hardware \textit{p-}circuit implementations, lateral shifts of the \textit{p-}bit response had to be eliminated by adjusting synaptic biases to calibrate the experiment \cite{borders_integer_2019,pervaiz_probabilistic_2019}. By contrast in this demonstration, since the biases are learned during operation, no calibration phase is necessary. This is a significant advantage since learning can account for transistor and \textit{s-}MTJ variations between \textit{p-}bits. After obtaining the response of all \textit{p-}bits, the learning experiment is performed (see methods section \ref{subsec: experiment} for more detail about the experimental procedure). \\
The goal of the learning process is that the \textit{p-}bits fluctuate according to a set data distribution. Since at every point in time the \textit{p-}bits can just be in one bipolar state, to monitor the training progress, the distribution of the \textit{p-}bits $P_{\rm Exp} (t)$ is collected as a histogram of the \textit{p-}bit output states $([m_1,m_2,m_3,m_4,m_5]+1)/2=[A,B,C_{\rm in},S,C_{\rm out}]$ over a fixed time window of 60 s, normalized to 1 and compared to the ideal distribution of a full adder given by the 8 lines of the truth table (see Table \ref{table: FA}). The experimental distribution at $t = 0$, $P_{\rm Exp} (t = 0)$ is shown in Fig. \ref{fig:FA} (b). At the start of learning the weights and biases are small and the distribution is close to a uniform random distribution. However, due to slight mismatches in the \textit{p-}bit response of every individual \textit{p-}bit [Fig. \ref{fig:FA} (a)] some peaks are more prominent than others. The distribution at the end of learning $P_{\rm Exp} (t = 3000s)$ is shown in Fig. \ref{fig:FA} (c), where the highest peaks correspond to the correct distribution for the FA, demonstrating the circuit’s ability to learn the given data distribution. To compare two probability distributions quantitatively the Kullback–Leibler divergence (KL-divergence) defined by $KL(P_\mathrm{Ideal} || P_\mathrm{Exp}(t))=\sum_\mathbf{\mathrm{m}} P_\mathrm{Ideal}(\mathbf{\mathrm{m}}) \log(P_\mathrm{Ideal}(\mathbf{\mathrm{m}})/P_\mathrm{Exp}(\mathbf{\mathrm{m}},t))$ is commonly used \cite{kullback_information_1951}. Fig. \ref{fig:FA} (d) shows the learning performance measured by the KL divergence versus time $t$. The difference between the ideal data distribution and the experimental distribution decreases significantly in the first 500 s of learning. At the end of learning the KL divergence reaches a value of around 0.2. We note that as long as the learned peaks are about equal, the KL divergence can be reduced further by increasing all weight values equally i.e. decreasing the temperature of the Boltzmann machine.\\
In Fig. \ref{fig:weights}, the 10 weights voltages across the capacitors $V_{\rm i,j}  = V_{\rm v;i,j}  -V_{\rm C;i,j}$ extracted from the circuit are shown. The weights are measured throughout the whole learning process. The blue lines show the weight voltages for the ideal MTJ. After around 500 s the weights saturate and do not change anymore. In the supplementary material \cite{Note2}, the weights values are compared to the weight matrix commonly used for the FA in logic applications \cite{hassan_voltage-driven_2019,pervaiz_weighted_2019}.\\

\begin{center}
\begin{table}
 \begin{tabular}{|c | c | c | c | c | c | c|} \hline
\begin{tabular}{@{}c@{}} $A$ \\ $v'_1$\end{tabular} & \begin{tabular}{@{}c@{}} $B$ \\ $v'_2$\end{tabular} & \begin{tabular}{@{}c@{}} $C_\mathrm{in}$ \\ $v'_3$\end{tabular} & \begin{tabular}{@{}c@{}} $S$ \\ $v'_4$\end{tabular} & \begin{tabular}{@{}c@{}} $C_\mathrm{out}$ \\ $v'_5$\end{tabular} & $P_\mathrm{Ideal}(v)$\\ \hline\hline
 0 & 0 & 0 & 0 & 0 & 0.125 \\  \hline
 0 & 0 & 1 & 1 & 0 & 0.125 \\  \hline
 0 & 1 & 0 & 1 & 0 & 0.125 \\  \hline
 0 & 1 & 1 & 0 & 1 & 0.125 \\  \hline
 1 & 0 & 0 & 1 & 0 & 0.125\\   \hline
 1 & 0 & 1 & 0 & 1 & 0.125 \\  \hline  
  1 & 1 & 0 & 0 & 1 & 0.125 \\  \hline 
 1 & 1 & 1 & 1 & 1 & 0.125 \\  \hline
\end{tabular}
 \caption{\textbf{Truth Table of Full Adder:} $A$ and $B$ are inputs, $C_{\rm in}$ is the carry in, $S$ the sum and $C_{\rm out}$ the carry out. In the Boltzmann machine context, all visible units are equivalent so that inputs and outputs can be written as $v'_{1-5}$. The bipolar training vectors $v_{\rm i}$ of Eq.\eqref{eq:learning_rule} can be calculated from the truth table by converting them from binary to bipolar $v_{\rm i}=2v'_{\rm i}-1$ where $[v'_1,v'_2,v'_3,v'_4,v'_5]=[A,B,C_{\rm in},S,C_{\rm out}]$ for the data distribution. $P_\mathrm{Ideal}(v)$ is the ideal data probability distribution where every line has a probability of $p=1/8=0.125$.}
\label{table: FA}
\end{table}
\end{center} 

\subsection{Full Adder learning with nonideal MTJ}
\begin{figure*}
    \setlength\abovecaptionskip{-0.5\baselineskip}
    \centering
    \includegraphics[width=1\linewidth]{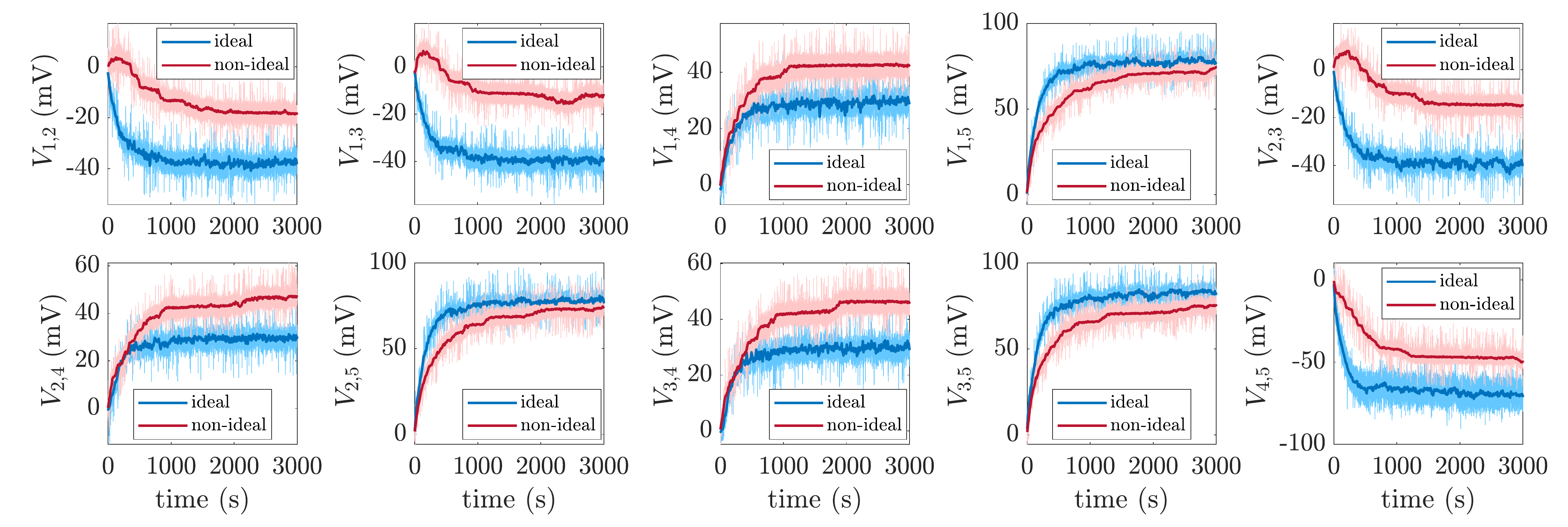} 
    \vspace{0.1in}
    \caption{\textbf{Weight voltages during FA learning:} The 10 weight voltages are shown during the 3000 s of learning. Blue lines are the weights learned with the ideal MTJ circuit; red lines show the weights for the nonideal MTJ circuit. The solid lines in the middle are the moving average of the actual weights taken over a window of 10 s.}
    \label{fig:weights}
\end{figure*}

To examine the effects of variability, we investigate the learning experiment implemented with fabricated \textit{s-}MTJs (see methods section \ref{subsec: fabrication} for more details regarding the fabrication). Fig. \ref{fig:FA} (e) shows the $V_{\rm OUT}$ versus $V_{ \rm IN}$ characteristics for the 5 MTJ-based \textit{p-}bits averaged over 15 s. At the transition point between the stochastic and the deterministic region of the response curve, the slope of the response is sharper compared to the center of the curve, which shows a gradual increase. The combination of these two characteristics leads to a nonideal \textit{p-}bit response that deviates from the ideal response described by Eq.\eqref{eq:binary_stochastic_neuron}. The reason for the distorted shape of the \textit{p-}bit response is due to the fact that the MTJs show stochastic behavior for a large window of current flow in the order of $> 10 \ \mu \mathrm{A}$. The change of the current flow in the MTJ/transistor branch due to change voltage at the gate of the transistor is not large enough to pin the MTJ to $R_{\rm P}$ or $R_{\rm AP}$ state. This leads to the distorted shape of the \textit{p-}bit response in Fig. \ref{fig:FA} (e). For best MTJ characteristics, the stochastic range for current flow should be in the order of around 5 $\mu$A in the design used here.\\
Fig. \ref{fig:FA} (f) and (g) show the histogram of $P_{\rm Exp}$ during the first and last 60 s of learning. At the end of learning the 8 desired peaks are the largest, showing that even though the learning algorithm is based on an ideal \textit{p-}bit response derived from the Boltzmann distribution, the circuit can still learn the desired functionality. Despite the noted nonidealities, the KL divergence saturates to a level comparable between ideal and nonideal MTJ as shown in Fig. \ref{fig:FA} (d). This can be explained by the fact that \textit{in situ} learning has the capabilities to counter device-to-device variations by adjusting weights and biases to fit the system (see supplementary information \cite{Note2} for more details on the learned bias voltages). \\
In Fig. \ref{fig:weights}, the red lines show the weight voltages of the nonideal MTJ over the duration of the learning process. It can be clearly seen that the weights differ significantly between the ideal and nonideal \textit{p-}bit implementation while achieving similar performance in the KL-divergence, leading to the conclusion that feedback in the system between data and \textit{p-}bit outputs is able to learn around variations, a crucial ingredient to achieve a high level of performance under device variability. In the supplementary information \cite{Note2} a system simulation on the MNIST dataset \cite{lecun_mnist_2010} is presented to show that the variation tolerance exists when the proposed circuit is scaled up.\\
The fact that the circuit can learn around variations can be useful not just for classical machine learning tasks like classification or unsupervised learning but also for tasks that have been demonstrated on probabilistic computers like optimization \cite{borders_integer_2019,sutton_intrinsic_2017}, inference \cite{faria_hardware_2021,faria_implementing_2018} or invertible logic \cite{lv_experimental_2019,camsari_stochastic_2017}. Instead of externally setting the coupling between \textit{p-}bits, an additional learning task could improve the performance of the \textit{p-}circuit by assuring that the coupling between the \textit{p-}bits is adjusted to the exact hardware \textit{p-}bit response.  In addition, the proposed hardware can be used to represent many different distinct probability distributions by adjusting the coupling between \textit{p-}bits accordingly.
For the particular combination of MTJ and transistor, voltage change at the input can change the output of the \textit{p-}bit on a transistor response time scale. Because the transistor response can be faster than the implemented synapse, for this particular experiment each \textit{p-}bit is updated sequentially through the microcontroller instead of autonomously to preserve functionality (see Ref. \cite{sutton_autonomous_2020} for more details).\\

\subsection{Weight extraction}
\begin{figure}
    \setlength\abovecaptionskip{-0.5\baselineskip}
    \centering
    \includegraphics[width=1\linewidth]{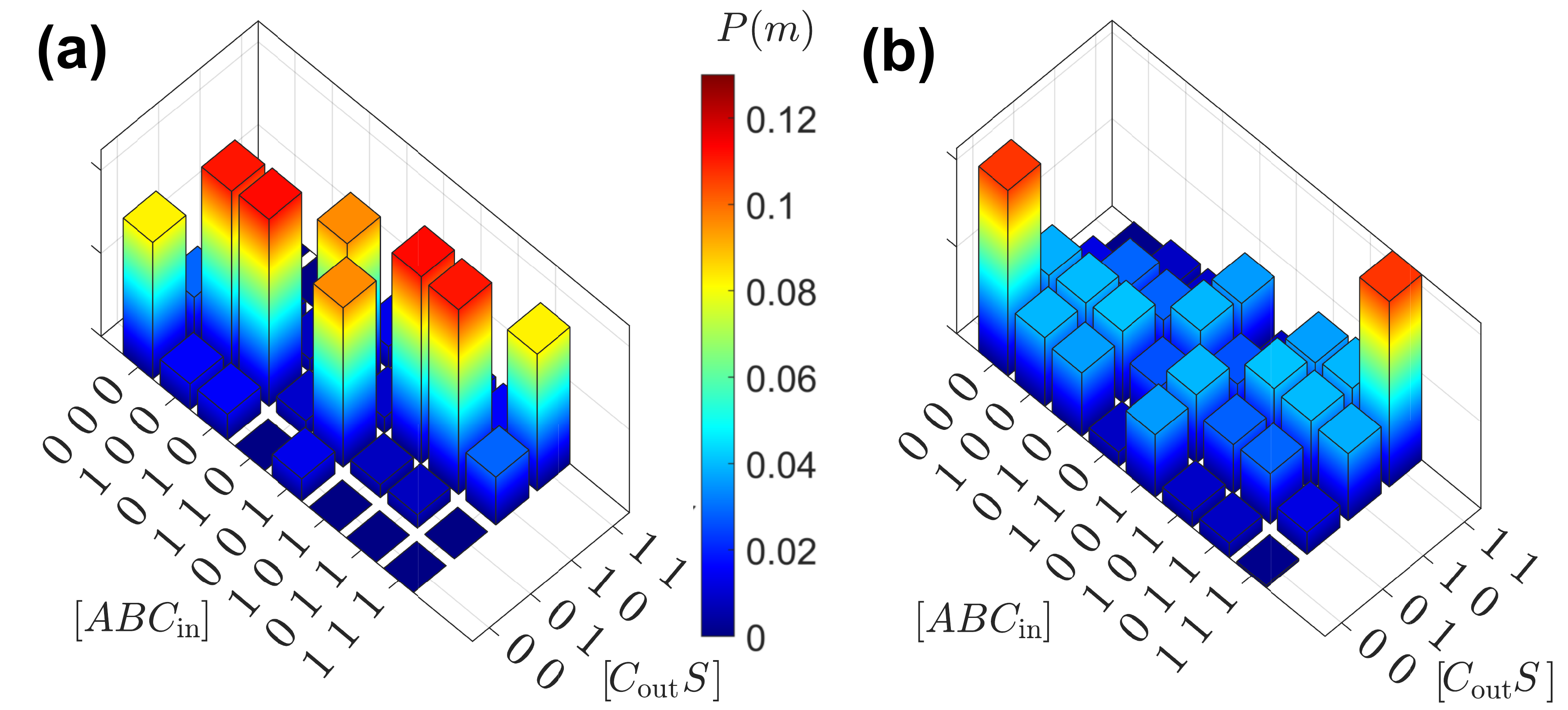} 
    \vspace{0.1in}
    \caption{\textbf{Boltzmann distribution obtained from learned weights:} (a) Boltzmann distribution $P(m)=1/Z \exp(-\beta E)$ with energy $E=-\sum W_{\rm i,j} m_{\rm i} m_{\rm j}$ computed by using the learned weights $W_{\rm i,j}$ of the FA with the emulated ideal \textit{s-}MTJ \textit{p-}bit circuit where the bipolar \textit{p-}bit states $([m_1,m_2,m_3,m_4,m_5]+1)/2=[A,B,C_{\rm in},S,C_{\rm out}]$. (b) Boltzmann distribution $P(m)$ computed by using the learned weights of the FA with the nonideal \textit{s-}MTJ \textit{p-}bit circuit. Biases are set to 0.}
    \label{fig:Boltzmann}
\end{figure}

In the previous sections, we compare the distribution of the output configurations of the hardware \textit{p-}bits averaged over 60 s with the ideal distribution by taking the Kullback-Leibler divergence. In this section we compare how the weights extracted as voltages across the capacitors in the circuit would perform on an ideal platform i.e. to the Boltzmann distribution where $P(m) \propto \exp{\big[-\beta E(m)\big]}$] and $\beta$ is the inverse temperature of the system. The temperature in a Boltzmann machine is a constant factor that all weights and biases are multiplied with and represents how strongly coupled the \textit{p-}bits are with each other. The comparison has particular relevance since the nonideal effects during learning should have an effect on the weights compared to the weights that would be learned on an ideal machine. Fig. \ref{fig:Boltzmann} shows the Boltzmann distribution with the weights of Fig. \ref{fig:weights}. The conversion factor between the voltages V across the capacitors and dimensionless weights W of the Boltzmann distribution represented by the temperature factor $\beta$ is chosen in a way that the relative difference between the peaks of the distribution can be seen clearly. To reduce the effect of noise, the weight values are averaged over the last 10 s of learning. For the example of the FA, it is known from the truth table that an ideal system has no bias. Hence, we do not use the extracted bias but set it to 0 for the Boltzmann distribution. In Fig. \ref{fig:Boltzmann} (a) it can be clearly seen that compared to Fig. \ref{fig:FA} (c) the learned distribution differs more from the ideal distribution since the peaks are not as uniform. The peaks for configuration $[ABC_{\rm in}]$ = 000, $[C_{\rm out} S]$ = 00 and $[ABC_{\rm in}]$ = 111, $[C_{\rm out} S]$ = 11 are not as prominent as the other 6 peaks that have been learned. This discrepancy becomes even more visible in Fig. \ref{fig:Boltzmann} (b) compared to Fig. \ref{fig:FA} (g) where the weights used in the Boltzmann distribution are learned using a less ideal response of the \textit{p-}bits. Here, only peaks $[ABC_{\rm in}]$ = 000,$[C_{\rm out} S]$ = 00 and $[ABC_{\rm in}]$ = 111,$[C_{\rm out} S]$ = 11 are prominent. This shows that the learned weights fit to the activation of the hardware \textit{p-}bits but not for the ideal Boltzmann distribution. Hence, we can conclude that the probabilistic computer adapted to the nonideal \textit{p-}bit response during the \textit{in situ} learning process.\\
The results presented in this section suggest that learning and inference must be performed on the same hardware to operate reliably. In contrast, initially training on this nonideal machine, then transferring the weight values to an ideal system to complete convergence and perform the programmed task could allow for a hardware-based speed-up of the typically time-consuming weight training step. This is similar in spirit to using pretrained weights in a neural network \cite{hinton_better_2012,he_rethinking_2019}. While this can be a disadvantage, the advantages of using the efficient and compact learning circuit that can be used for training and inference should outweigh the problems of transferability between platforms.\\
In this section, we show that device-to-device variations can be countered by performing hardware aware \textit{in situ} learning by comparing the learning performance of two systems, one system with ideal \textit{p-}bit responses and the other with nonideal \textit{p-}bit responses that differ significantly compared to Eq.\eqref{eq:binary_stochastic_neuron}. We show that the overall performance is the same for both systems after the training is finished while the learned weights (Fig. \ref{fig:weights}) are different. However, we also show that if the weights are extracted from the learning circuit and used to calculate the Boltzmann distribution, the obtained distribution differs substantially from the desired data distribution [Fig. \ref{fig:Boltzmann} (b)]. These observations show clearly that the circuit can learn around device-to-device variations. \\

\section{Discussion}
In this paper, we present a proof-of-concept demonstration of an autonomously operating fully connected Boltzmann machine using MTJ-based \textit{p-}bits. Furthermore, we show how device-to-device variations can be countered by performing hardware aware \textit{in situ} learning. In the following paragraphs, we compare the presented probabilistic computer with other platforms like conventional CMOS architectures.\\
On the device level, the closest digital CMOS alternative to the MTJ-based \textit{p-}bit is a linear feedback shift register (LFSR), without considering the analog tunability of the \textit{p-}bit. A detailed comparison between \textit{p-}bit versus LFSR has been performed by Borders et al.\cite{borders_integer_2019}. The compact MTJ-based \textit{p-}bit uses around 10x less energy per random bit and has about 300x less area than a 32-bit LFSR. Besides these advantages, a standard LFSR is not tunable like the hardware \textit{p-}bit and relies on pseudo randomness. The \textit{p-}bit based on an \textit{s-}MTJ relies on thermal noise and is, hence, a true random number generator. This can be significant for applications for which the quality of the randomness is relevant. \\
On the system level, the \textit{p-}bits in combination with the synapse (Eqs.\eqref{eq:binary_stochastic_neuron} and \eqref{eq:synaptic_function}) are utilized to collect samples of the distribution given by the current weights to update the weights according to the correct gradient. Collecting statistics by sampling drives the learning process since every sample is directly utilized to update the weight voltages (Eq.\eqref{eq:learning_rule}). Thus, the numbers of samples per unit time are significant for the speed of the learning process. The MTJ fluctuation time of the \textit{p-}bit $\tau_N$ is a significant time scale for the generation of samples since it describes how fast Eq.\eqref{eq:binary_stochastic_neuron} can be computed in hardware. The learning time constant $\tau_L$ has to be larger than the MTJ fluctuation time $\tau_N$ to collect enough statistics to ensure convergence of the learning process. To ensure that every \textit{p-}bit input is correctly calculated based on the state of the other \textit{p-}bits, it is necessary that the synapse time $\tau_S$ is smaller than $\tau_N$. In this experiment, since the synapse time defined by the microcontroller is in the order of 100 $\mu$s to 1 ms, $\tau_N$ is in the order of 10 - 100 ms which results in slow training in the order of $10^3$ s. However, it has to be noted that the time scales of the circuit can be reduced significantly in an integrated version of the proposed circuit where the synapse based on crossbar architectures can operate with GHz speeds with execution times down to 10 ps \cite{sutton_autonomous_2020,peng_gu_technological_2015,cai_power-efficient_2020} and the fluctuation time of \textit{s-}MTJs can be in the order of 100 ps \cite{kaiser_subnanosecond_2019,hassan_low-barrier_2019,pufall_large-angle_2004}. This would allow a substantial decrease of $\tau_L$ and an increase of the learning speed by up to 9 orders of magnitude. Regarding energy consumption of the synapse block, the efficient \textit{p-}bit building block presented here can be combined with any synapse option that provides the most power efficiency. For full inference operation, the RC array used here to represent weights as voltages requires a constant memory refresh similar to mainstream dynamic random-access memory (DRAM). To save energy during the learning process, the presented \textit{p-}bit building block could be combined with nonvolatile synapse implementations like memristive crossbar arrays \cite{ernoult_using_2019,bojnordi_memristive_2016,cai_power-efficient_2020}. The learned weights could also be extracted from the RC array and stored in a nonvolatile memory array after the learning process. 

The overall power consumption can be estimated using numbers from the literature. The MTJ-based \textit{p-}bit consumes about 20 $\mu$W \cite{hassan_low-barrier_2019}. In a memristive crossbar, each memristor consumes about 1 $\mu$W and operational amplifiers around 3 $\mu$W \cite{sutton_autonomous_2020,cai_power-efficient_2020,li_memristor-based_2013}. The XNOR operation consumes 10 $\mu$W. For the overall circuit with 5 \textit{p-}bits, 15 XNOR-gates and memristors, and 5 operational amplifiers would take approximately 300 $\mu$W. This is the projected power consumption of a fully-connected Boltzmann machine hardware shown in this work. For specified applications where less weight connections between neurons are needed (for example restricted Boltzmann machines in digital computers), the number of components can be reduced which results in improved power consumption. In this regard, the estimated power consumption in our work can also be significantly reduced by employing a higher-level approach.

Another significant advantage of the probabilistic circuit is that due to the compactness and area savings of the \textit{p-}bit, when scaling up, many more \textit{p-}bits can be put on a chip compared to CMOS alternatives like LFSRs. In addition, the \textit{p-}bit hardware implementation does not rely on any clocking in order to function and is hence autonomously operating. This has the advantage that many autonomously operating \textit{p-}bits can function in parallel leading to an overall acceleration of the operation. In this context, it has to be noted that the information of the current state of a \textit{p-}bit has to be propagated to all other \textit{p-}bits that are connected to it on a time scale $\tau_S$  that is much shorter than the neuron time $\tau_N$ for the probabilistic circuit to function properly. When the \textit{p-}bit fluctuation time varies between different \textit{p-}bit it has to be assured that the fastest \textit{p-}bit with fluctuation time $\tau_{N,f}$  fluctuates slower than $\tau_S$. Depending on the sparsity of the weight matrix and the ratio of $\tau_S$ to $\tau_N$, the number of parallel operating \textit{p-}bits has to be adjusted to ensure fidelity of the operation \cite{sutton_autonomous_2020}. In a recent paper by Sutton et al. \cite{sutton_autonomous_2020} an FPGA design was implemented that emulates a probabilistic circuit where the MTJ based \textit{p-}bit is envisioned as a drop-in replacement. In this complete system-level hardware realization of a \textit{p-}computer that can perform only inference not learning, a drastic reduction in area footprint of the compact \textit{p-}bit design compared to digital implementations is confirmed. This shows that an integrated version of the proposed learning circuit based on the \textit{p-}computer architecture could be very beneficial. 

While we address that device-to-device variations of the shape and shift of the \textit{p-}bit response can be accounted for by hardware-aware learning, it is worthwhile to note that rate variation of the stochastic MTJ between \textit{p-}bits cannot be reduced by this approach. The system will in the worst case learn as fast as the fluctuation rate of the slowest \textit{p-}bit $\tau_{N,s}$ which can slow down the overall operation. However, in the case of \textit{p-}bits with stochastic MTJs where the thermal barrier of the magnet in the free layer is in the order of $k_B T$, the fluctuation rate does not go exponentially with the size of the magnet making the system less susceptible to rate variations \cite{brown_thermal_1963,coffey_thermal_2012,kaiser_subnanosecond_2019,hassan_low-barrier_2019}. It has to be noted that a way to reduce rate variation in probabilistic circuits based on stable MTJs that are biased using voltages and magnetic fields has been presented by Lv et al.\cite{lv_experimental_2019}.

We note that the fluctuation rate will also be affected by the temperature of the probabilistic circuit. When increasing the temperature, the fluctuation rate of the \textit{p-}bits will increase exponentially. However, the temperature variation will not affect the average \textit{p-}bit response of the MTJ. For proper operation it has to be assured that the synapse time $\tau_S$ is shorter than the fluctuation time $\tau_{N,f}$ of the fastest fluctuating \textit{p-}bit. As overall design criteria for the autonomous circuit the following conditions have to be met: $\tau_S \ll \tau_{N,f}$ and $\tau_{N,s}\ll \tau_L$.\\
In conclusion, we show a proof-of-concept demonstration of a fully connected probabilistic computer built with MTJ-based \textit{p-}bits that can perform learning. We present multiple learning examples for up to 5 \textit{p-}bits and 15 learning parameters. The learning is robust and can operate even with strong device-to-device variations due to hardware-aware learning. This shows that when scaled up and with faster fluctuating building blocks, probabilistic computers could accelerate computation while reducing energy cost for a wide variety of tasks in the machine learning field such as generative learning or sampling, as well as for tasks that could benefit from variation tolerance like optimization or invertible logic. 

\section{Materials and Methods}
\label{methods}
\subsection{MTJ fabrication \& Characterization}
\label{subsec: fabrication}

The MTJs used in this work are fabricated with a stack structure as follows, from the substrate side: Ta(5)/ Pt(5)/ [Co(0.4)/Pt(0.4)]\textsubscript{6}/ Co(0.4)/ Ru(0.4)/ [Co(0.4)/Pt(0.4)]\textsubscript{2}/ Co(0.4)/ Ta(0.2)/ (Co\textsubscript{0.25}Fe\textsubscript{0.75})\textsubscript{75}B\textsubscript{25}(1)/ MgO/ (Co\textsubscript{0.25}Fe\textsubscript{0.75})\textsubscript{75}B\textsubscript{25}(1.7)/ Ta(5)/ Ru(5)/ Ta(50). The numbers in parentheses are the nominal thicknesses in nanometers. All films are deposited on a thermally oxidized silicon substrate by dc and rf magnetron sputtering at room temperature. The stacks are then processed into circular MTJs with nominal junction size of 20-25 nm in diameter by electron beam lithography and argon ion milling. The samples are annealed at 300$^\circ$C in vacuum for an hour. MTJs are then cut out from wafers and bonded with wires to IC sockets to be placed in the \textit{p-}bit circuit board. To determine nonideal MTJs with suitable characteristics, the MTJ resistance is measured by sweeping the current from negative to positive values, and the time-averaged and high-frequency signals are read across a voltmeter and oscilloscope, respectively. We measure an approximate tunnel magnetoresistance ratio of 65\% fluctuating between an average $R_{\rm P} = 18 \ {\rm k \Omega}$ and $R_{\rm AP} = 30 \ {\rm k \Omega}$. The current at which the resistance switches by half is determined to be $I_{50/50}$, which is the bias current at which the MTJs will spend equal time in the P and AP states. The $I_{50/50}$ used in this work ranges from 3 to 5 $\mu$A. We measure the average fluctuation time $\tau_N$ by performing retention time measurements when the MTJ is in either the high (AP) or the low (P) state using voltage readings from the oscilloscope. To ensure reliable collection of data, the oscilloscope sampling rate is set ten times faster than the fastest recorded fluctuation time of the MTJ. The retention times used in this work range from 1 ms to 100 ms.

\subsection{Hardware implementation of the \textit{p-}bit}
\label{subsec: p-bit}

\begin{figure}
    \setlength\abovecaptionskip{-0.5\baselineskip}
    \centering
    \includegraphics[width=1\linewidth]{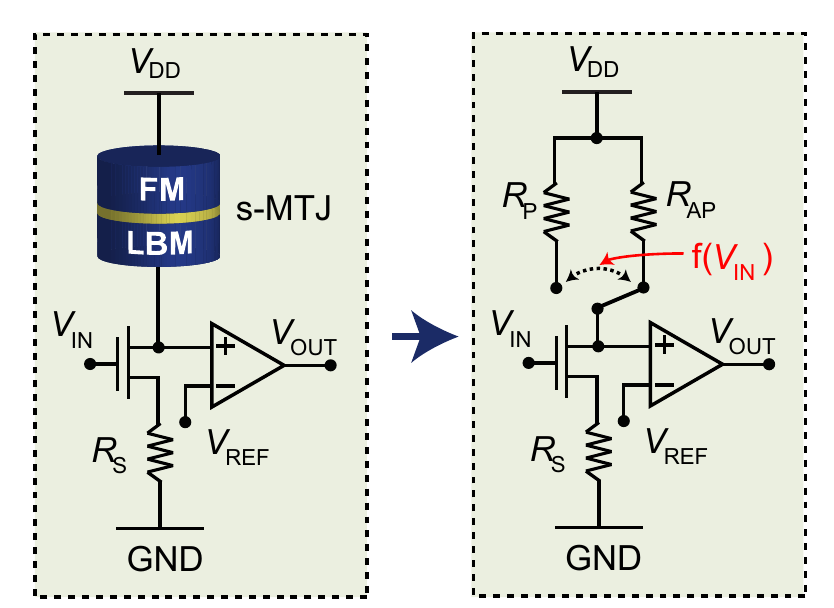} 
    \vspace{0.1in}
    \caption{\textbf{Multiplexer emulation:} The \textit{s-}MTJ based \textit{p-}bit on the left is modeled by a multiplexer that switches randomly between $R_{\rm P}$ and $R_{\rm AP}$ but as a function of $V_{\rm IN}$ so that the right statistics are preserved \cite{pervaiz_probabilistic_2019}.}
    \label{fig:MUXmodel}
\end{figure}

Eq.\eqref{eq:binary_stochastic_neuron} is implemented with the \textit{s-}MTJ based \textit{p-}bit proposed by Camsari et al.\cite{camsari_implementing_2017} and experimentally demonstrated by Borders et al.\cite{borders_integer_2019}. The \textit{p-}bit implementation in this paper follows Ref. \cite{borders_integer_2019} and is built with an \textit{s-}MTJ in series to a transistor (2N7000,T0-92-3 package) and a source resistor $R_{\rm S}$. The supply voltage of the MTJ transistor branch is set to $V_{\rm DD}=200$ mV whereas the remaining circuit operates at $V_{\rm DD}=5$ V. The source resistance $R_{\rm S}$ is chosen so that $I_{50/50}$ is flowing through the circuit when $V_{\rm IN}=1.95$ V. The transistor is biased in the subthreshold region. The voltage at the drain of the transistor is then thresholded using a comparator (AD8694, 16-SOIC package) with a bandwidth of 10 MHz. The reference voltage is chosen to be $V_{\rm REF}=V_{\rm DD}-I_{50/50}\big(\frac{R_{\rm P}+R_{\rm AP}}{2}\big)$. We have used a comparator to add another node where we can fine tune $V_{\rm REF}$. However, in an integrated circuit the transistor should be chosen so that $V_{\rm REF}=V_{\rm DD}/2$ so that the comparator can be replaced by a simple inverter as simulated in references \cite{camsari_implementing_2017,hassan_low-barrier_2019,kaiser_probabilistic_2020}. The overall \textit{p-}bit is then just built with 1 MTJ and 3 transistors.  For the experiment with ideal MTJs, the \textit{s-}MTJ is emulated by a multiplexer (MUX) model that includes all major characteristics of a real \textit{s-}MTJ and has been developed by Pervaiz et al.\cite{pervaiz_probabilistic_2019} as illustrated in Fig. \ref{fig:MUXmodel}. The \textit{s-}MTJ is emulated by providing a noise signal to the MUX where the statistics of the noise depend on $V_{\rm IN}$ and are generated using a microcontroller that switches between a resistor $R_{\rm P}$ and $R_{ \rm AP}$ representing the two resistive states of the \textit{s-}MTJ. Here, the resistors values are chosen to be $R_{\rm P}=11 \ {\rm k \Omega}$ and $R_{\rm AP}=22 \ {\rm k \Omega}$. The advantage of this approach is that the MTJ parameters like stochastic range and resistance can be easily manipulated in this model. For the MUX, a MAX 394 quad analog multiplexer is used.

\subsection{Implementation of the synapse}
\label{subsec: synapse}

The synapse is implemented with an Arduino MEGA microcontroller and an 8-channel PMOD DA4 Digital-Analog-Converter. The digital output voltages of the \textit{p-}bits $\{V_{\rm OUT}\}$ are fed into the microcontroller together with the analog weight voltages $\{V_{\rm C}\}$ of the learning circuit. The internal Analog-Digital-Converter (ADC) of the microcontroller is used for sensing the weight voltages. Eq.\eqref{eq:synaptic_function} is then computed and the analog input voltages $\{V_{\rm IN}\}$ are wired back to the neurons by utilizing the DAC. To reduce the synapse time in every iteration of the synapse operation, only one of the 15 analog voltages are read out and updated. This does not affect the circuit performance since the capacitor voltages $V_C$ are changing slowly. The synapse operation time $\tau_S$ is $< 1$ ms which is shorter than the MTJ fluctuation time. The condition $\tau_S \ll \tau_N$ has to be satisfied to ensure fidelity of the autonomous operation of the \textit{p-}circuit.

\subsection{Implementation of weight updating}
\label{subsec: weights}

For proper operation it is important that the learning time constant $\tau_L$ is much larger than the neuron time $\tau_N$. To achieve this, a high RC constant is chosen with a 1 M$\Omega$ resistor and a $10 \ \mu$F capacitor. Since this circuit has a high resistance in series to the capacitor, to ensure that the reading of the weight voltage does not discharge the capacitor, a buffer stage is used between the capacitor and the synapse. The buffer is implemented with an operational amplifier (AD8694, 16-SOIC package).

For the FA experiment, the voltage gain factor $A_v$ of Eq.\eqref{eq:learning_circuit} is chosen to be 3 which turned out to be a reasonable value for achieving a good degree of regularization while achieving high peaks in the learned distribution. The voltage gain operation is performed with the microcontroller. Additional details regarding Eq.\eqref{eq:learning_circuit} can be found in Ref. \cite{kaiser_probabilistic_2020}. 

For learning the correlations $m_{\rm i} m_{\rm j}$, represented by voltage $V_{\rm m;i,j}$, are crucial. To obtain the current correlations between neuron $m_{\rm i}$ and $m_{\rm j}$ their product has to be computed. This is done here by using another microcontroller. Since the output $m$ is bipolar ($m \in \{-1,1\}$) only negative or positive correlation is possible. Voltage $V_{\rm m;i,j}$ is limited by the output voltages of the DAC which has a range from 0 V to 2.5 V. $V_{\rm m;i,j}$ can hence be calculated by solving $V_{\rm m;i,j}= (m_{\rm i}m_{\rm j}+1)/2 \cdot 2.5$ V. Voltage $V_{\rm m;i,j}$ is fed back to the corresponding RC element by utilizing another DAC. The described operation is the same as computing the XNOR operation between two binary variables. Hence, the operation is straight forward and the programmability of the microcontroller not essential for operation of the circuit.

\subsection{Experimental procedure}
\label{subsec: experiment}

Before the start of training the capacitor is fully discharged so that $V_{\rm i,j}(t=0)=0$ V corresponding to $V_{\rm C;i,j}(t=0)=V_{\rm v;i,j}$. At $t=0$ the training starts and voltages $\{V_{C}\}$ and the \textit{p-}bit output voltages $\{V_{\rm OUT}\}$ are measured at sampling frequency $f_S$. The training is run for $T=3000s$. \\

The data is collected with an \textit{NI USB-6351 X SERIES DAQ} that has analog inputs for the 15 weights and biases and digital inputs for the 5 \textit{p-}bit outputs. The software \textit{Labview} is utilized to record data with a sampling frequency of $f_S=1$ kHz.\\

In this paper we have trained the bias due to mismatch of \textit{p-}bit responses together with the bias needed to learn the data distribution. In principle, these can be separated to obtain a better bias value that can be used on other platforms. However, this separation of calibration and learning is only possible for the bias of every \textit{p-}bit and not for the weights connecting them since the calibration cannot be performed with ideal \textit{p-}bit responses with the hardware system.\\

\subsection{Mapping of the truth table to node voltages for learning}
\label{subsec: mapping}
For a fully visible Boltzmann machine with $N$ neurons, $(N+1)N/2$ weights and biases have to be learned. The goal for learning is that the fully trained network has the same distribution as the data distribution. For a FA, the data distribution is given by the truth table shown in table 1.
The data distribution can be described by a matrix in which the number of columns is equal to the number of neurons $N$ and the number of rows is equal to the number of training examples d. For the biases, another neuron unit with value 1 is added so that there are $(N + 1)$ columns. For the example of a FA, $N = 5$ and $d = 8$ for 8 lines in the truth table. The matrix $V_{\rm FA}$ is then a 6x8 matrix where all 0s of the truth table are converted to -1s since we are using the bipolar representation:
\begin{equation}
V_{\rm FA}=
\begin{bmatrix}
-1 & -1 & -1 & -1 & -1 & 1 \\
-1 & -1 & 1 & 1 & -1 & 1 \\
-1 & 1 & -1 & 1 & -1 & 1 \\
-1 & 1 & 1 & -1 & 1 & 1 \\
1 & -1 & -1 & 1 & -1 & 1 \\
1 & -1 & 1 & -1 & 1 & 1 \\
1 & 1 & -1 & -1 & 1 & 1 \\
1 & 1 & 1 & 1 & 1 & 1 \\
\end{bmatrix}
\label{eq: VFA}
\end{equation}
The density matrix is then calculated by computing $D= V^TV/d$ which is a 6x6 matrix for the FA:
\begin{equation}
D_{\rm FA}=\frac{V_{\rm FA}^T V_{\rm FA}}{d}=
\begin{bmatrix}
1 & 0 & 0 & 0 & 0.5 & 0\\
0 & 1 & 0 & 0 & 0.5 & 0\\
0 & 0 & 1 & 0 & 0.5 & 0\\
0 & 0 & 0 & 1 & -0.5 & 0\\
0.5 & 0.5 & 0.5 & -0.5 & 1 & 0\\
0 & 0 & 0 & 0 & 0 & 1\\

\end{bmatrix}
\label{eq: DFA}
\end{equation}
with $d=8$. The values in the last column of the density matrix correspond to the average value of every neuron in the data distribution and are used to learn the biases. Only the terms above the diagonal of $D$ are needed and converted to voltages $V_{\rm v;i,j}$ in the circuit. Since the DAC operates with positive voltages in the range of $0 \ \mathrm{V}$ to $2.5 \ \mathrm{V}$, $V_{\rm v;i,j}=(D_{\rm i,j}+1)/2 \cdot 2.5$ V.\\

\acknowledgements
J.K. thanks A.Z. Pervaiz for helpful discussions. This work was supported in part by ASCENT, one of six centers in JUMP, a Semiconductor Research Corporation (SRC) program sponsored by DARPA and in part by JST-CREST JPMJCR19K3, JSPS Kakenhi 19J12206, and Cooperative Research Projects of RIEC. K.Y.C gratefully acknowledges support from Center for Science of Information (CSoI), an NSF Science and Technology Center, under grant CCF-0939370.

\bibliography{library}

\end{document}

\renewcommand{\thetable}{S\arabic{table}}  
\renewcommand{\thefigure}{S\arabic{figure}} 
\renewcommand{\theequation}{S\arabic{equation}} 

\title{Supplemental Material: Hardware-aware \textit{in situ} learning based on stochastic magnetic tunnel junctions}
\author{Jan Kaiser}
\affiliation{Department of Electrical and Computer Engineering, Purdue University, West Lafayette, IN, 47906 USA}
\author{William A. Borders}
\affiliation{Laboratory for Nanoelectronics and Spintronics, Research Institute of Electrical Communication, Tohoku University, Sendai, Japan}
\author{Kerem Y. Camsari}
\email{camsari@ece.ucsb.edu}
\affiliation{Department of Electrical and Computer Engineering, University of California, Santa Barbara, CA, 93106 USA}
\author{Shunsuke Fukami}
\email{s-fukami@riec.tohoku.ac.jp}
\affiliation{Laboratory for Nanoelectronics and Spintronics, Research Institute of Electrical Communication, Tohoku University, Sendai, Japan}
\affiliation{Center for Innovative Integrated Electronic Systems, Tohoku University, Sendai, Japan.}
\affiliation{Center for Spintronics Research Network, Tohoku University, Sendai, Japan.}
\affiliation{Center for Science and Innovation in Spintronics, Tohoku University, Sendai, Japan.}
\affiliation{WPI-Advanced Institute for Materials Research, Tohoku University, Sendai, Japan.}
\author{Hideo Ohno}
\affiliation{Laboratory for Nanoelectronics and Spintronics, Research Institute of Electrical Communication, Tohoku University, Sendai, Japan}
\affiliation{Center for Innovative Integrated Electronic Systems, Tohoku University, Sendai, Japan.}
\affiliation{Center for Spintronics Research Network, Tohoku University, Sendai, Japan.}
\affiliation{Center for Science and Innovation in Spintronics, Tohoku University, Sendai, Japan.}
\affiliation{WPI-Advanced Institute for Materials Research, Tohoku University, Sendai, Japan.}
\author{Supriyo Datta}
\affiliation{Department of Electrical and Computer Engineering, Purdue University, West Lafayette, IN, 47906 USA}
\date{\today}

\pacs{}
\maketitle
\section{Learned weights and biases}
In the main manuscript, the learned probability distribution of the full adder is analyzed. In this section the actual weight and bias voltages across the capacitors compared to the ideal FA weights. The weight matrix for a FA for an ideal p-computer with ideal sigmoidal p-bit responses is the following and has been part of several works \cite{hassan_voltage-driven_2019,pervaiz_weighted_2019}:
\begin{equation}
W_{\rm FA}=
\begin{bmatrix}
0 & -1 & -1 & 1 & 2 \\
-1 & 0 & -1 & 1 & 2 \\
-1 & -1 & 0 & 1 & 2 \\
1 & 1 & 1 & 0 & -2  \\
2 & 2 & 2 & -2 & 0 \\
\end{bmatrix}
\label{eq: FA_matrix}
\end{equation}
Since the ideal FA probability distribution is symmetric, the bias vector is 0 and can be disregarded here.
In Fig. 3 of the main manuscript the weights voltages across the capacitors $V_{\rm i,j}  = V_{\rm v;i,j}  -V_{\rm C;i,j}$ extracted from the RC-circuit are shown. Since the \textit{p-}bit response has units of voltage whereas the ideal \textit{p-}bit response is unitless, there is a constant conversion factor between the $W_{\rm FA}$ and the weight voltages in Fig. 3. Since the \textit{p-}bit responses differ for both cases, the learned weights voltages are not identical. The weights are learned to fit to the given non-ideal response of each \textit{p-}bit.

However, it can be clearly seen that the general structure of the different weight voltage matrix extracted from the experiment and $W_{\rm FA}$ is similar at the end of the learning process. For example $-2 V_{\rm 1,2} \approx -2 V_{\rm 1,3} \approx 2 V_{\rm 1,4} \approx V_{\rm 1,5}$ which corresponds to $-2 W_{1,2} = -2 W_{1,3} = 2 W_{1,4} = W_{1,5}$ in Eq.\eqref{eq: FA_matrix}. This makes the point that even though the weights learned in this experiment are not ideal due to the non-ideal \textit{p-}bit responses, they are related to the weights of an ideal \textit{p-}computer. Initializing with the weights learned on a hardware probabilistic computer could hence reduce learning time when trying to learn based on an ideal Boltzmann distribution as mentioned in the main manuscript.

\begin{figure}[t]
    \setlength\abovecaptionskip{-0.5\baselineskip}
    \centering
    \includegraphics[width=1\linewidth]{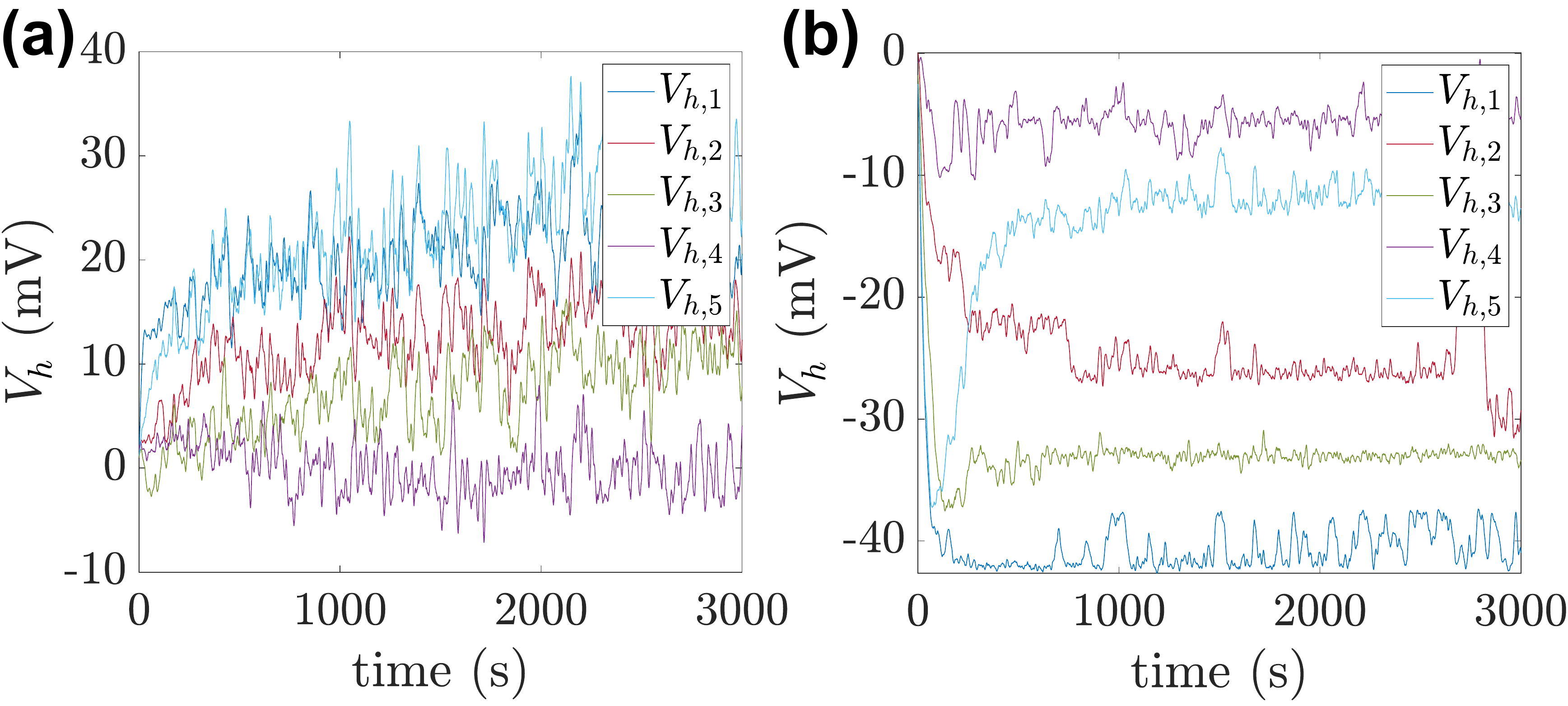} 
    \vspace{0.1in}
    \caption{\textbf{Bias voltages during FA learning:} (a) Bias voltages $\{V_h\}$ across the capacitors for ideal MTJ, (b) Bias voltages $\{V_h\}$ for non-ideal MTJ.}
    \label{fig:biases}
\end{figure}

In Fig. \ref{fig:biases} the learned biases are shown. Since the ideal learned biases are 0, the biases learned in this experiment account for the shifted \textit{p-}bit responses away from the ideal response center at $V_{\rm IN}=1.95 \ \mathrm{V}$. Since the \textit{p-}bit responses for the non-ideal MTJ in Fig. 3 (a) of the main manuscript are shifted to the left all biases are negative and bigger than the biases needed for the emulated ideal MTJ.

\section{Boltzmann machine learning algorithm}
\label{section: Logliklihood}
For learning probability distributions in the context of energy-based models like Boltzmann machines the common learning algorithm is gradient ascent of the log-likelihood given by
\begin{equation}
L(W;V)=\frac{1}{N} \sum_V \log \frac{1}{Z} \exp[-E(v_n;W)]
\label{eqn: log-liklihood}
\end{equation}
where $Z$ is the partition function and the data distribution is given by $V=\{v_n\}_{n=1}^d$ \cite{carreira-perpinan_contrastive_2005, koller_probabilistic_2009}. Here, the inverse temperature is set to $\beta=1$. The gradient ascent update rule is given by
\begin{equation}
W_{\rm i,j}(t+1)=W_{\rm i,j}(t)+\epsilon \frac{\partial L(W;V)}{\partial W}\Big|_{W(t)}
\end{equation}
with the learning rate $\epsilon$. Solving the derivative of $L(W;V)$ gives \cite{carreira-perpinan_contrastive_2005}
\begin{equation}
W_{\rm i,j}(t+1)=W_{\rm i,j}(t)-\biggl\langle \frac{\partial E(m)}{\partial W}\biggl\rangle_{\rm data}+\biggl\langle\frac{\partial E(m)}{\partial W}\biggl\rangle_{\rm model}
\label{eq: generalLR}
\end{equation}
The data-term in the derivative evolves from $\exp(-E(v_n;W)$ and the model-term from the partition function $Z$ in Eq.\eqref{eqn: log-liklihood}. With energy given by $E(m)=-\sum W_{\rm i,j} m_{\rm i} m_{\rm j}$, the Boltzmann machine learning rule is
\begin{equation}
W_{\rm i,j}(t+1)=W_{\rm i,j}(t)+\epsilon \big(\langle v_{\rm i}v_{\rm j} \rangle- \langle m_{\rm i}m_{\rm j} \rangle\big)
\label{eq: BLR} 
\end{equation}
Eq.(3) of the main manuscript is the time-continuous version of Eq.\eqref{eq: BLR} where the averaged correlation $\langle m_{\rm i}m_{\rm j} \rangle$ is replaced with the sampled correlation $m_{\rm i}m_{\rm j}$ (compare Ref. \cite{kaiser_probabilistic_2020}).

It has to be noted that the learning rule in Eq.\eqref{eq: BLR} assumes ideal sigmoidal \textit{p-}bit responses since it is derived from Boltzmann law. However, in this paper the same learning rule is also applied when \textit{p-}bit responses are non-ideal and good learning results are achieved.

\section{Learning of AND, OR and XOR gate}
In this section learning examples with smaller numbers of \textit{p-}bits are presented. The same PCB is used but only 3 \textit{p-}bits and 6 RC elements are used for the AND and OR gate and 4 \textit{p-}bits and 10 RC elements are used for the XOR gate. Here, the ideal MUX model is used.

\label{section: ANDORXOR}
\subsection{Learning of an AND-Gate}
For an AND-Gate the truth table matrix in the bipolar representation $V$ with an added column with $+1$ for the bias is given by
\begin{equation}
V_{\rm AND}=
\begin{bmatrix}
-1 & -1 & -1 & 1 \\
-1 & 1 & -1 & 1 \\
1 & -1 & -1 & 1 \\
1 & 1 & 1 & 1
\end{bmatrix}
\end{equation}

\begin{figure}[t]
    \setlength\abovecaptionskip{-0.5\baselineskip}
    \centering
    \includegraphics[width=1\linewidth]{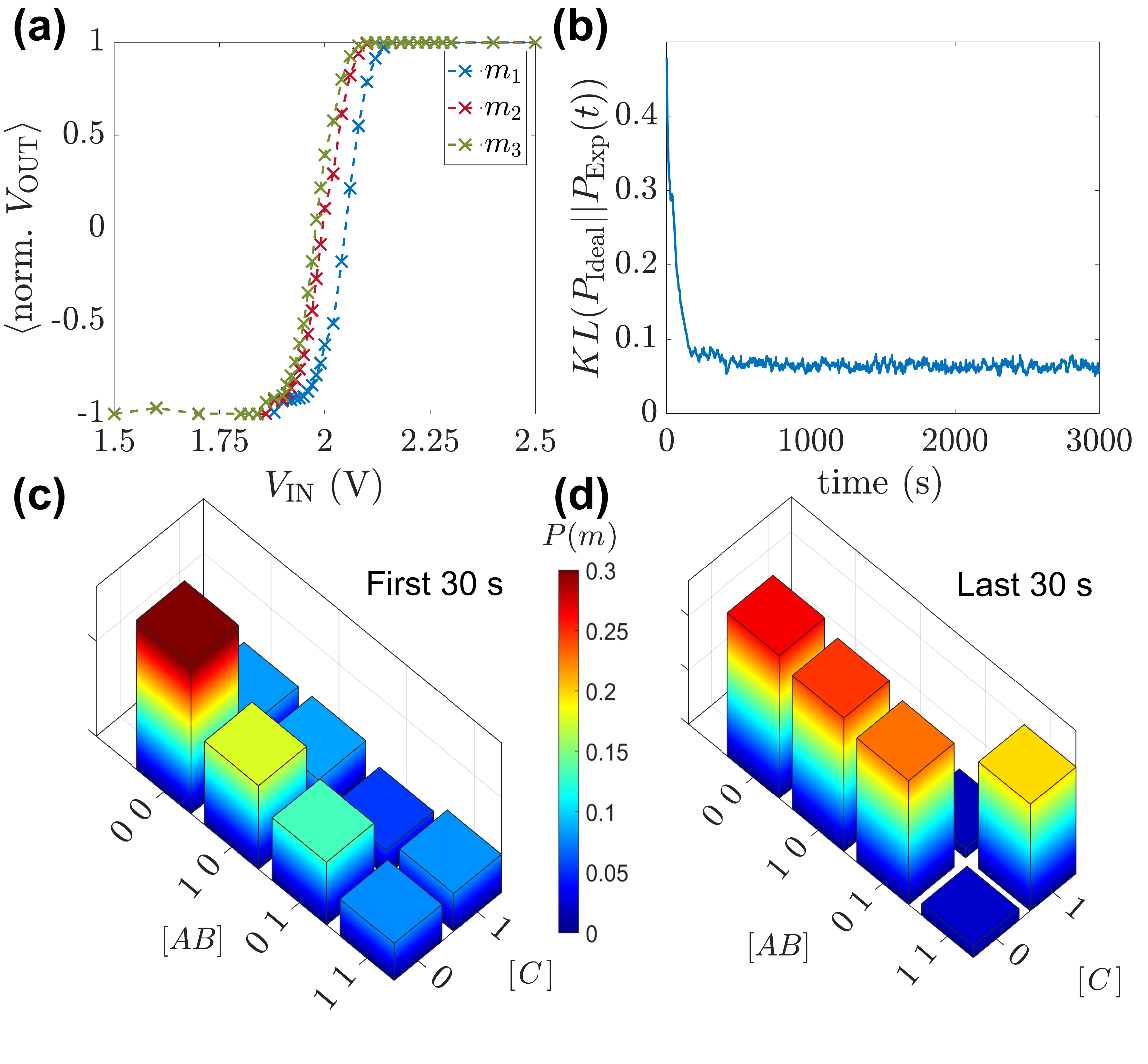} 
    \vspace{0.1in}
    \caption{\textbf{AND-Gate:} (a) Average response for the 3 \textit{p-}bits used in the AND-Gate with average normalized output voltage $\langle \mathrm{norm.} \ V_{\rm OUT,i} \rangle = 2\cdot \langle V_{\rm OUT,i} \rangle /{V_{\rm DD} -1=\langle m_{\rm i} \rangle}$. (b) KL-divergence between ideal $P_{\rm Ideal}(m)$ and experimental distribution $P_{\rm Exp}(m)$ with \textit{p-}bit output states $([m_1,m_2,m_3]+1)/2=[A,B,C]$ where $m_{\rm i}=2 \cdot V_{\rm OUT,i}/V_{\rm DD} -1$ collected as a histogram is plotted against time. The experimental distribution is obtained over 30 s of learning. (c) Experimental distribution of emulated ideal MTJ circuit $P_{\rm Exp}(m)$ over for the first 30 s of learning. (d) Experimental distribution over the last 30 s of learning. The voltage gain $A_v$ is set to 3.}
    \label{fig:results_AND}
\end{figure}

The density matrix is then given by 
\begin{equation}
D_{\rm AND}=\frac{V_{\rm AND}^T V_{\rm AND}}{d}=
\begin{bmatrix}
1 & 0 & 0.5 & 0 \\
0 & 1 & 0.5 & 0 \\
0.5 & 0.5 & 1 & -0.5 \\
0 & 0 & -0.5 & 1
\end{bmatrix}
\end{equation}
with $d=4$. In total 6 parameters have to be learned. Fig. \ref{fig:results_AND} (a) shows the \textit{p-}bit response of the 3 \textit{p-}bits used for AND gate learning. Fig. \ref{fig:results_AND} (b) shows the KL divergence and Fig. \ref{fig:results_AND} (c) and (d) show the histogram at the start and at the end of learning. For the AND gate after learning out of the 8 possible configurations, the 4 desired states become most and equally probable.

\subsection{Learning of an OR-Gate}
For an OR-Gate the truth table matrix in the bipolar representation $V$ with an added column with $+1$ for the bias is given by
\begin{equation}
V_{\rm OR}=
\begin{bmatrix}
-1 & -1 & -1 & 1 \\
-1 & 1 & 1 & 1 \\
1 & -1 & 1 & 1 \\
1 & 1 & 1 & 1
\end{bmatrix}
\end{equation}

\begin{figure}[t]
    \setlength\abovecaptionskip{-0.5\baselineskip}
    \centering
    \includegraphics[width=1\linewidth]{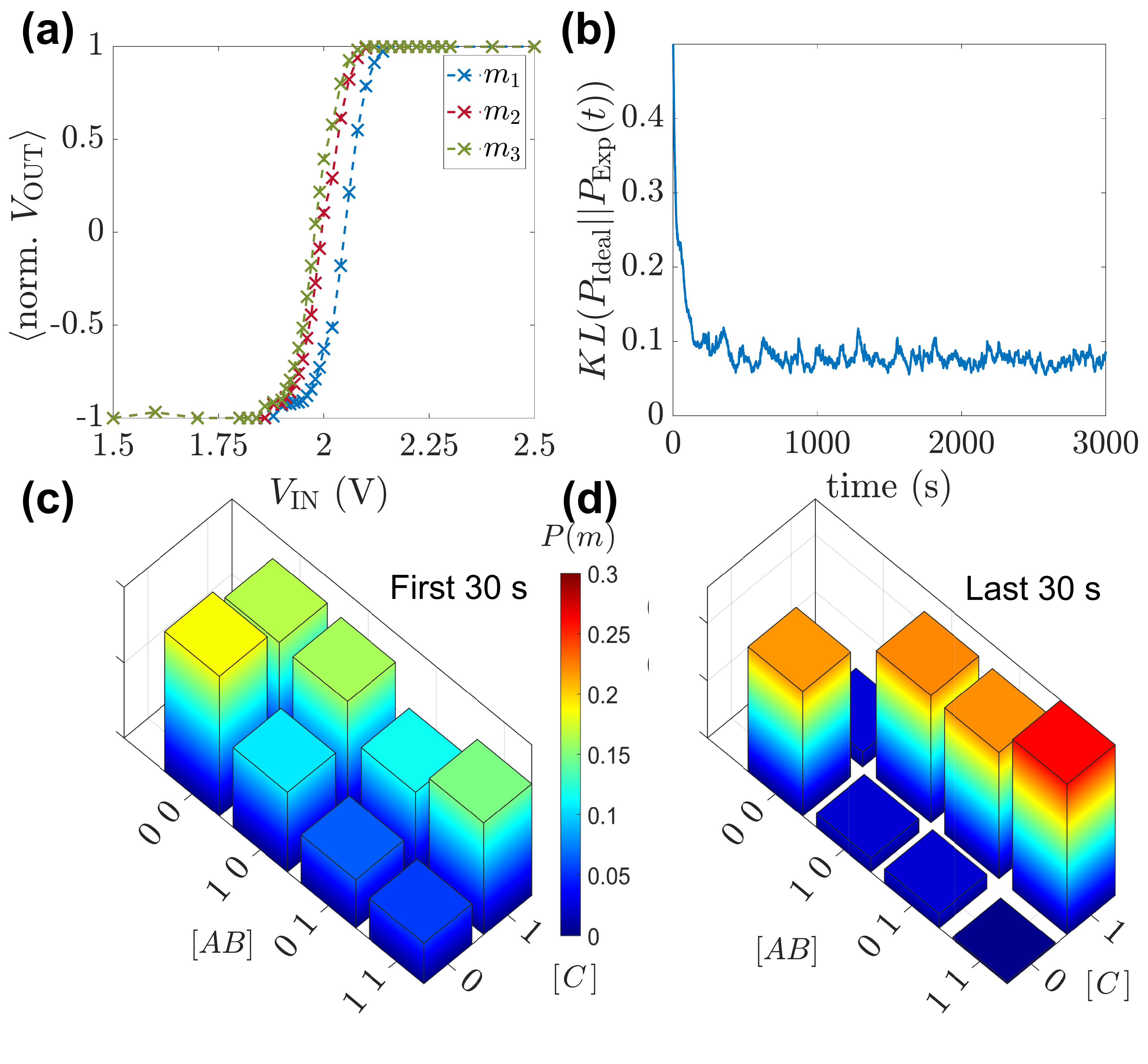} 
    \vspace{0.1in}
    \caption{\textbf{OR-GATE:} (a) Average response for the 3 \textit{p-}bits used in the AND-Gate with average normalized output voltage $\langle \mathrm{norm.} \ V_{\rm OUT,i} \rangle = 2\cdot \langle V_{\rm OUT,i} \rangle /{V_{\rm DD} -1=\langle m_{\rm i} \rangle}$. (b) KL-divergence between ideal $P_{\rm Ideal}(m)$ and experimental distribution $P_{\rm Exp}(m)$ with \textit{p-}bit output states $([m_1,m_2,m_3]+1)/2=[A,B,C]$ where $m_{\rm i}=2 \cdot V_{\rm OUT,i}/V_{\rm DD} -1$ collected as a histogram is plotted against time. The experimental distribution is obtained over 30 s of learning. (c) Experimental distribution of emulated ideal MTJ circuit $P_{\rm Exp}(m)$ over for the first 30 s of learning. (d) Experimental distribution over the last 30 s of learning. The voltage gain $A_v$ is set to 3.}
    \label{fig:results_OR}
\end{figure}

The density matrix is then given by 
\begin{equation}
D_{\rm OR}=\frac{V_{\rm OR}^T V_{\rm OR}}{d}=
\begin{bmatrix}
1 & 0 & 0.5 & 0 \\
0 & 1 & 0.5 & 0 \\
0.5 & 0.5 & 1 & 0.5 \\
0 & 0 & 0.5 & 1
\end{bmatrix}
\end{equation}
with $d=4$. In total 6 parameters have to be learned. Fig. \ref{fig:results_OR} (a) shows the \textit{p-}bit response of the 3 \textit{p-}bits used for OR gate learning. Fig. \ref{fig:results_OR} (b) shows the KL divergence and Fig. \ref{fig:results_OR} (c) and (d) show the histogram at the start and at the end of learning. For the OR gate after learning out of the 8 possible configurations, the 4 desired states become most and equally probable.

\subsection{Learning of an XOR-Gate}
For an XOR-Gate even though there are just 2 inputs and 1 output an auxiliary neuron is needed to be able to learn the XOR functionality. Without an additional \textit{p-}bit, all entries of the density matrix are 0 which corresponds to no learning at all. Here, we choose the auxiliary neuron to be in the first column of the $V_{\rm XOR}$ matrix and to be 1 for the first entry and -1 for the last 3 entries of the XOR truth table matrix. 
\begin{equation}
V_{\rm XOR}=
\begin{bmatrix}
1 & -1 & -1 & -1 & 1 \\
-1 & -1 & 1 & 1 & 1 \\
-1 & 1 & -1 & 1 & 1 \\
-1 & 1 & 1 & -1 & 1 \\
\end{bmatrix}
\end{equation}

\begin{figure}
    \setlength\abovecaptionskip{-0.5\baselineskip}
    \centering
    \includegraphics[width=1\linewidth]{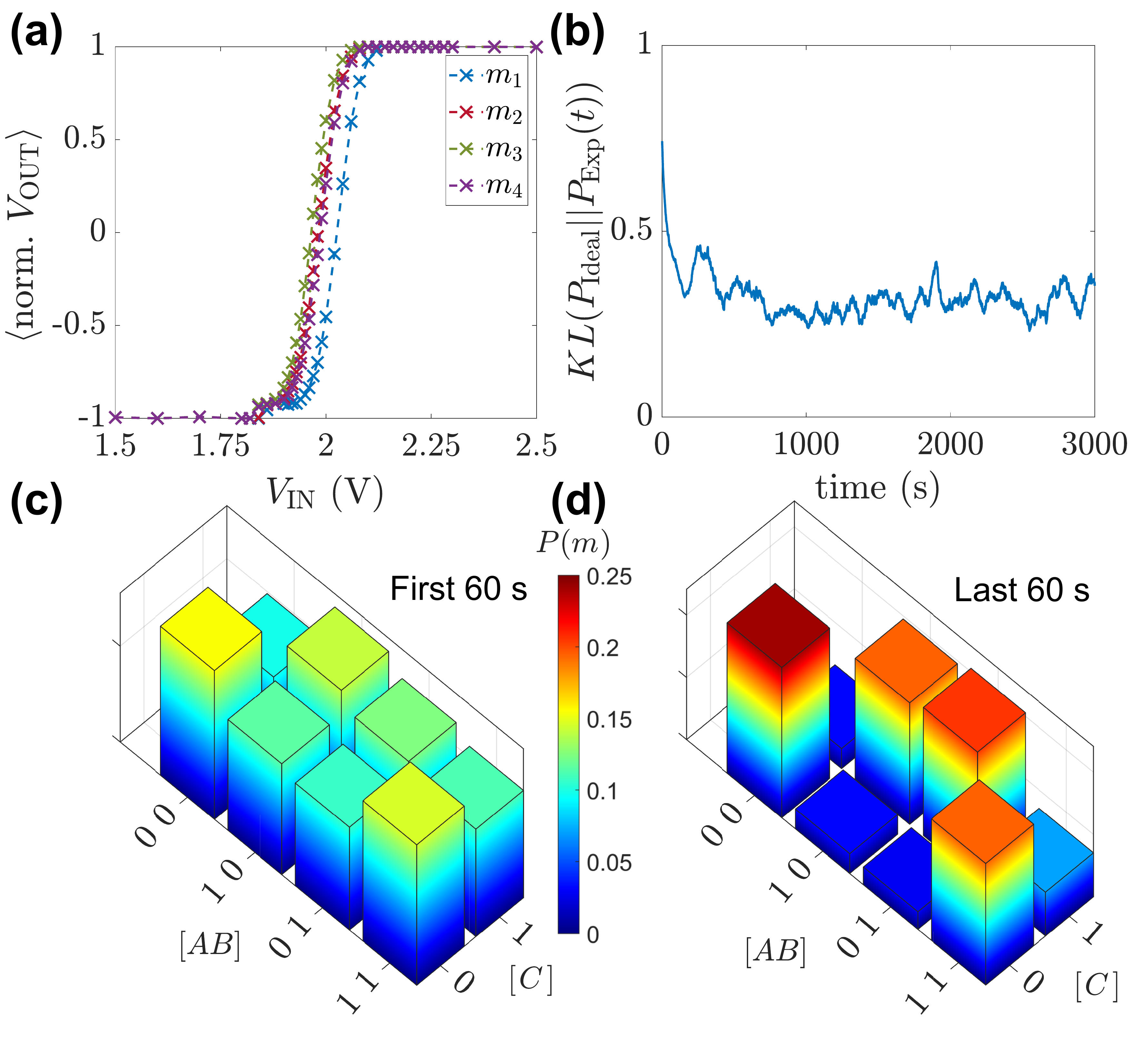} 
    \vspace{0.1in}
    \caption{\textbf{XOR-Gate:}(a) Average response for the 4 \textit{p-}bits used in the AND-Gate with average normalized output voltage $\langle \mathrm{norm.} \ V_{\rm OUT,i} \rangle = 2\cdot \langle V_{\rm OUT,i} \rangle /{V_{\rm DD} -1=\langle m_{\rm i} \rangle}$. (b) KL-divergence between ideal $P_{\rm Ideal}(m)$ and experimental distribution $P_{\rm Exp}(m)$ of \textit{p-}bit output states $([m_2,m_3,m_4]+1)/2=[A,B,C]$ where $m_{\rm i}=2 \cdot V_{\rm OUT,i}/V_{\rm DD} -1$ collected as a histogram is plotted against time. The experimental distribution is obtained over 60 s of learning. (c) Experimental distribution of emulated ideal MTJ circuit $P_{\rm Exp}(m)$ over for the first 60 s of learning. (d) Experimental distribution over the last 60 s of learning. The voltage gain $A_v$ is set to 4.}
    \label{fig:results_XOR}
\end{figure}

The density matrix is then given by 
\begin{equation}
D_{\rm XOR}=\frac{V_{\rm XOR}^TV_{\rm XOR}}{d}=
\begin{bmatrix}
1 & -0.5 & -0.5 & -0.5 & -0.5\\
-0.5 & 1 & 0 & 0 & 0\\
-0.5 & 0 & 1 & 0 & 0\\
-0.5 & 0 & 0 & 1 & 0\\
-0.5 & 0 & 0 & 0 & 1\\

\end{bmatrix}
\end{equation}
with $d=4$. It can be clearly seen that without the first column in $V_{\rm XOR}$ all off-diagonal terms of the $D_{\rm XOR}$ would be 0. In total 10 parameters have to be learned. Fig. \ref{fig:results_XOR} (a) shows the \textit{p-}bit response of the 4 \textit{p-}bits used for XOR gate learning. Fig. \ref{fig:results_XOR} b shows the KL divergence and Fig. \ref{fig:results_XOR} (d) show the histogram at the start and at the end of learning. The figure just shows the histogram of the 3 of the 4 \textit{p-}bits ($([m_2,m_3,m_4]+1)/2=[A,B,C]$) without plotting the states of the auxiliary \textit{p-}bit $m_1$. For the XOR gate after learning, the 4 desired states become most and equally probable out of the 8 possible configurations.
\begin{figure}
    \setlength\abovecaptionskip{-0.5\baselineskip}
    \centering
    \includegraphics[width=1\linewidth]{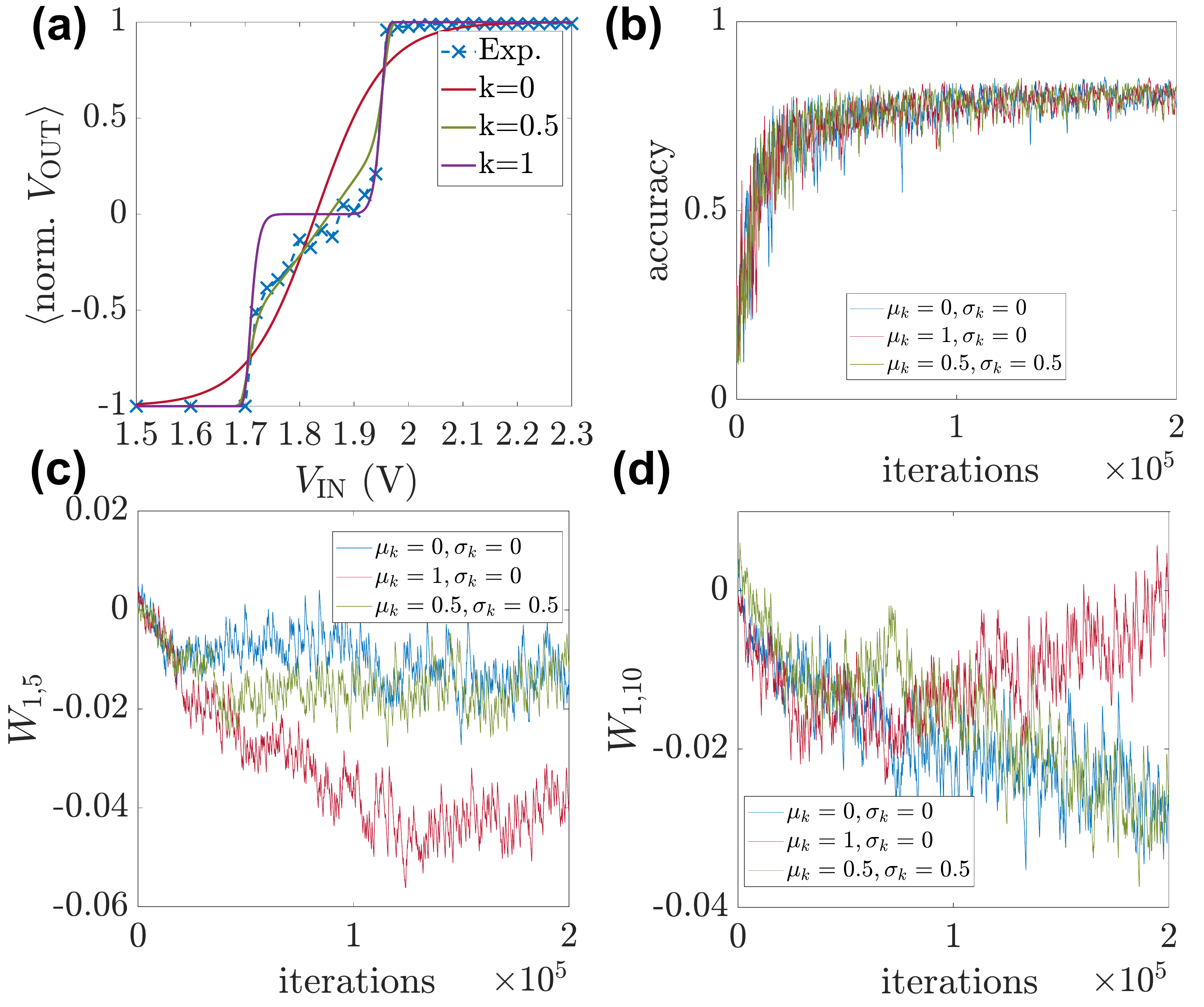} 
    \vspace{0.1in}
    \caption{\textbf{Learning with behavioral \textit{p-}bit model \cite{faria_hardware_2021} on MNIST dataset:}  (a) Experimental \textit{p-}bit response is compared to the model of Eq.\eqref{eq:act} for different values of $k$ where $x$ is fitted to the input voltage $V_{ \rm IN}$. (b) Test set accuracy on the MNIST dataset during training (c), (d) Example weights during training. Following parameters are used in the behavioral model: neuron time $\tau_N=150$ ps, synapse time $\tau_S=10$ ps, transistor time $\tau_T=25$ ps and $\Delta t=1$ ps. The used learning parameters are $\epsilon=10^{-5}$, $\lambda$=0.0125 here.}
    \label{fig:MNIST}
\end{figure}

\begin{algorithm}[H]
\SetAlgoLined
 Given a data set $X$, calculate density matrix $D=X X^T$\;
 Initialize $W$ to 0 and $m$ randomly\;
 \For{t=0:T (number of iterations)}{
	Get $m$ from \textit{p-}bit sampling procedure (Eqs. 1,2)\; 
	Calculate $M=mm^T$\;
	Update $W_{\rm i,j}=W_{\rm i,j}+\epsilon(D_{\rm i,j}-M_{\rm i,j}-\lambda W_{\rm i,j})$ (Eq. 3)\;
	Set diagonal terms of $W$ to 0\;
  }
 \caption{Behavioral model of proposed learning circuit.}
 \label{algorithm}
\end{algorithm}

\section{Simulations of the proposed circuit for larger networks}
In this section we use a behavioral model on the MNIST dataset \cite{lecun_mnist_2010} to show that the variation tolerance observed in our proof-of-concept experiment can be transferred to larger scale. It has to be noted that the implemented circuit in our proof-of-concept experiment is a fully visible Boltzmann machine that does not make use of any hidden neurons. This means that the states of all nodes of the Boltzmann machine are given by the data distribution. Hidden neurons add representational power to a Boltzmann machine and are needed for reaching high absolute accuracy on image recognition tasks like MNIST\cite{le_roux_representational_2008}.
The MNIST dataset has 60000 training images and 10000 test images with 28x28 pixels with digits from 0 to 9. The fully visible Boltzmann network used here consists of 794 \textit{p-}bits (28x28=784 + 10 \textit{p-}bits used as labels).  The MNIST dataset is transformed into bipolar values and Algorithm \ref{algorithm} which emulates the circuit's behavior is used for learning. For every iteration of the \textit{p-}bit update procedure, the behavioral model proposed by Faria et al.\cite{faria_hardware_2021} for the hardware \textit{p-}bit implementation is utilized, a model that has been benchmarked against SPICE simulations. In addition, the activation function is changed to account for device-to-device variations.
To model the behavior of the proposed circuit we use the formula 
\begin{equation}
act(x,k)=\tanh[(1-k) \cdot x+k \cdot x^{11}]
\label{eq:act}
\end{equation}

where $k \in [0,1]$ parameterizes how ideal the response of the \textit{p-}bit is.
In Fig. \ref{fig:MNIST} (a), Eq.\eqref{eq:act} is compared to a non-ideal \textit{p-}bit response observed in the experiment. For $k=0$ the ideal \textit{p-}bit response is observed whereas for $k=1$ the \textit{p-}bit response looks like a staircase. It can be clearly seen that the model is very close to the observed experimental behavior of the \textit{p-}bits. To simulate the variation behavior, the factor k is drawn from a Gaussian distribution with mean $\mu_k$ and standard deviation $\sigma_k$ for every \textit{p-}bit.
In Fig. \ref{fig:MNIST} (b) the accuracy of the circuit is shown for every iteration of Algorithm \ref{algorithm} for different distributions of k for each \textit{p-}bit. To obtain test results, the 784 \textit{p-}bits that correspond to the pixels are clamped to the bipolar test data and the label \textit{p-}bits are fluctuating freely. The \textit{p-}bit with the highest probability of being '1' is used for the classified digit. The learning is performed for different values of $\mu_k$ and $\sigma_k$. After around $10^5$ iterations the accuracy saturates to about 81\% for all 3 curves shown while the learned weights differ [Fig. \ref{fig:MNIST} (c),d)]. This shows that the circuit can account for non-ideal \textit{p-}bit responses by learning the correct weights. The learning can account for the non-ideal \textit{p-}bit responses and still obtain similar accuracy. The behavioral model simulation suggests that the learning duration of the task shown in Fig. \ref{fig:MNIST} can be around 100 ns with $\Delta t=1$ ps and $10^5$ iterations in an ideally optimized integrated circuit using MTJ based \textit{p-}bits. The 81\% accuracy is due to the chosen fully visible network structure without any hidden units. The low performance of this model is not due to the hardware components but due to the low representational power of the fully visible Boltzmann machine \cite{le_roux_representational_2008}. The same circuit with hidden nodes can be for example implemented by time sharing the \textit{p-}bit circuit for collecting data and model statistics but is out of the scope of this paper.

\bibliographystyle{apsrev4-2}
\balance\bibliography{library}